\documentclass[useAMS,usenatbib]{mn2e}

\def\kms{km\,s$^{-1}$}

\def\M{M$_{\odot}$}

\def\lam{$\lambda$}

\def\Ha{H$\alpha$}

 \def\Ni{$^{56}$Ni}
 \def\Co{$^{56}$Co}
 
 \def\Mej{$M_{\rm ej}$}
\def\Mcsm{$M_{\rm CSM}$}
\def\Mni{$M_{\rm Ni}$}
\def\Rph{$R_{\rm phot}$}

\usepackage{graphicx, subfigure}
\usepackage[authoryear]{natbib}
\usepackage{placeins}
\usepackage{amsmath}

\title[PESSTO super-luminous supernovae]
{Super-luminous supernovae from PESSTO}

\author[Nicholl et al.]{M. Nicholl$^{1}$\thanks{E-mail: mnicholl03@qub.ac.uk},
S. J. Smartt$^{1}$, A. Jerkstrand$^{1}$, C. Inserra$^{1}$, J. P. Anderson$^{2}$, C. Baltay$^{3}$, 
\newauthor{S. Benetti$^{4}$, T.-W. Chen$^{1}$, N. Elias-Rosa$^{4}$, U. Feindt$^{5}$, M. Fraser$^{6}$, A. Gal-Yam$^{7}$,}
\newauthor{E. Hadjiyska$^{3}$, D. A. Howell$^{8,9}$, R. Kotak$^{1}$, A. Lawrence$^{10}$, G. Leloudas$^{11,12}$,}
\newauthor{S. Margheim$^{13}$, S. Mattila$^{14}$, M. McCrum$^{1}$, R. McKinnon$^{3}$, A. Mead$^{10}$,}
\newauthor{P. Nugent$^{15,16}$, D. Rabinowitz$^{3}$, A. Rest$^{17}$, K. W. Smith$^{1}$, J. Sollerman$^{18}$,}
\newauthor{M. Sullivan$^{19}$, F. Taddia$^{18}$, S. Valenti$^{8,9}$, E. S. Walker$^{3}$, D. R. Young$^{1}$}
\\
$^1$  Astrophysics Research Centre, School of Mathematics and Physics, Queen's University Belfast, 
Belfast, BT7 1NN,  UK \\
$^{2}$European Southern Observatory, Alonso de Cordova 3107, Vitacura, Casilla 19001, Santiago, Chile\\
$^{3}$Department of Physics, Yale University, New Haven, CT 06520-8121, USA\\
$^{4}$INAF - Osservatorio Astronomico di Padova, vicolo dell'Osservatorio 5, I-35122 Padova, Italy \\
$^{5}$Physikalisches Institut, Universit\"at Bonn, Nussallee 12, 53115 Bonn, Germany\\
$^{6}$Institute of Astronomy, University of Cambridge, Madingley Road, Cambridge, CB3 0HA, UK\\
$^{7}$Benoziyo Center for Astrophysics, Weizmann Institute of Science, Rehovot 76100, Israel\\
$^{8}$Department of Physics, University of California, Santa Barbara, Broida Hall, Mail Code 9530, Santa Barbara, CA 93106-9530, USA\\
$^{9}$Las Cumbres Observatory, Global Telescope Network, 6740 Cortona Drive Suite 102, Goleta, CA 93117, USA\\
$^{10}$University of Edinburgh, Institute for Astronomy, Royal Observatory, Blackford Hill, Edinburgh, EH9 3HJ, UK\\
$^{11}$The Oskar Klein Centre, Department of Physics, Stockholm University, 10691 Stockholm, Sweden \\
$^{12}$Dark Cosmology Centre, Niels Bohr Institute, University of Copenhagen, 2100 Copenhagen, Denmark \\
$^{13}$Gemini Observatory, Southern Operations Center, Casilla 603, La Serena, Chile\\
$^{14}$Finnish Centre for Astronomy with ESO (FINCA), University of Turku, V\"ais\"al\"antie 20, FI-21500 Piikki\"o, Finland\\
$^{15}$Department of Astronomy, University of California, Berkeley, B-20 Hearst Field Annex \#3411, Berkeley, CA 94720-3411, USA\\
$^{16}$Computational Cosmology Center, Computational Research Division, Lawrence Berkeley National Laboratory, 1 Cyclotron Road MS 50B-4206, Berkeley, CA 94720, USA\\
$^{17}$Space Telescope Science Institute, 3700 San Martin Dr., Baltimore, MD 21218, USA\\
$^{18}$The Oskar Klein Centre, Department of Astronomy, Stockholm University, 10691 Stockholm, Sweden \\
$^{19}$School of Physics and Astronomy, University of Southampton, Southampton, SO17 1BJ, UK\\
}

\begin{document}

\maketitle

\begin{abstract}

We present optical spectra and light curves for three hydrogen-poor super-luminous supernovae followed by the Public ESO Spectroscopic Survey of Transient Objects (PESSTO). Time series spectroscopy from a few days after maximum light to 100 days later shows them to be fairly typical of this class, with spectra dominated by Ca II, Mg II, Fe II and Si II, which evolve slowly over most of the post-peak photospheric phase. We determine bolometric light curves and apply simple fitting tools, based on the diffusion of energy input by magnetar spin-down, \Ni~decay, and collision of the ejecta with an opaque circumstellar shell. We investigate how the heterogeneous light curves of our sample (combined with others from the literature) can help to constrain the possible mechanisms behind these events. We have followed these events to beyond 100-200 days after peak, to disentangle host galaxy light from fading supernova flux and to differentiate between the models, which predict diverse behaviour at this phase. Models powered by radioactivity require unrealistic parameters to reproduce the observed light curves, as found by previous studies. Both magnetar heating and circumstellar interaction still appear to be viable candidates. A large diversity is emerging in observed tail-phase luminosities, with magnetar models failing in some cases to predict the rapid drop in flux. This would suggest either that magnetars are not responsible, or that the X-ray flux from the magnetar wind is not fully trapped. The light curve of one object shows a distinct re-brightening at around 100d after maximum light. We argue that this could result either from multiple shells of circumstellar material, or from a magnetar ionisation front breaking out of the ejecta.
\end{abstract}


%\clearpage

\section{Introduction} \label{intro}

In recent years, observational studies of supernovae (SNe) have been
revolutionized by a new generation of transient surveys,
which observe large areas of the sky without a bias for particular
galaxy types. The Palomar Transient Factory \citep[PTF;][]{rau2009},
Pan-STARRS1 \citep[PS1;][]{kai2010}, Catalina Real-Time Transient Survey
\citep[CRTS;][]{dra2009}, and La Silla QUEST \citep[LSQ;][]{balt2013}, for example, find
thousands of SNe per year, among which lurk some very unusual
objects. In particular, there has been much interest in a population
of very luminous blue transients inhabiting faint galaxies. \citet{qui2011}
were the first to obtain secure measurements of their redshifts, and establish these SNe as a new class,
through detections of narrow host galaxy Mg II \lam\lam~2796, 2803 absorption
lines. Typical redshifts $z \sim$ 0.2--0.5 implied absolute peak
magnitudes $M < -21$, making these SNe at least 10--100 times more luminous
than the usual thermonuclear (type Ia) and core-collapse
SNe. This intrinsic brightness enabled Pan-STARRS1 to extend the redshift
range to $z\sim1$ \citep{chom2011}. \citet{coo2012} have since detected
events as distant as $z\sim2$ and 4, in stacked images from the Canada-France-Hawaii Telescope Legacy Survey Deep Fields.

\citet{gal2012} reviewed these ``super-luminous'' supernovae (SLSNe),
and defined three sub-classes, roughly in analogy with conventional SN
nomenclature. The hydrogen-rich events were named SLSNe II. Many
of these show narrow and intermediate-width Balmer emission lines,
similar to normal SNe IIn, and
their light curves are likely powered by a collision between the SN ejecta and a massive, optically thick
circumstellar shell
\citep{smi2007,smi2008,mil2009,ben2014}. In this case, kinetic energy is thermalised in the opaque shell by radiation-dominated shocks \citep{che2011}, and diffuses out as observable light. The archetypal example of
this class is SN 2006gy \citep{smi2007b,ofe2007}.

The group identified by \citet{qui2011} were designated as SLSNe of Type I, since they are hydrogen-poor. The first examples were SN 2005ap
\citep{qui2007} and SCP-06F6 \citep{bar2009}. The early optical spectra of
these objects are dominated by broad absorptions of O II, and are very
blue and quite featureless around peak brightness. \citet{pas2010} showed that a relatively nearby object, SN 2010gx, evolved to spectroscopically resemble more typical SNe Ic, but with delayed line
formation relative to their normal-luminosity cousins \citep[also see][for late-time, and host galaxy, analysis]{chen2013}. The mechanism
that powers SLSN I or Ic remains undetermined. However, it is clear
that the light curves observed so far are incompatible
with models powered by the radioactive decay chain \Ni---\Co---$^{56}$Fe \citep{qui2011,gal2012,ins2013}, which is the usual energy source in
Type Ia or Ibc SNe. One plausible candidate is delayed heating by a central engine,
such as a spinning-down magnetar
\citep{woo2010,kas2010,des2012} or fall-back accretion
\citep{dex2012}, giving luminous light curves with power-law
declines. Another possibility is circumstellar interaction with
hydrogen-free shells \citep{woo2007}; however, it has not been shown whether
this can produce the observed spectra, which lack narrow lines. \citet{ins2013} collected extensive data on 5 SLSNe of Type Ic (in their nomenclature) at redshift $z<0.25$. They found that magnetar models could quantitatively reproduce the observed light curves.

The third sub-class is based on the decline rate of the SN luminosity. A small number of hydrogen-free SLSNe have light curve gradients after peak magnitude which are consistent with radioactive \Co~decay. This led \citet{gal2012} to propose a
classification name of SLSN-R. These may be the observational
counterparts of the long-predicted pair-instability supernovae
\citep[PISNe;][]{bar1967,rak1967}. In these models, photons in the cores of 130--250 \M~stars are sufficiently energetic
to decay into electron-positron pairs, and the conversion of
pressure-supporting radiation to rest-mass triggers contraction
followed by thermonuclear runaway. Only one published event, SN 2007bi \citep{gal2009,you2010}, has been considered a strong candidate
for a PISN. However, two very similar SLSNe, PTF12dam and PS1-11ap, with
better photometric and spectroscopic coverage, have since been shown by
\citet{nic2013} and \citet{mcc2013} 
to be inconsistent with PISN models, and
their early spectra resemble the other SLSNe Ic. Whether or not the fast (2005ap-like) and slowly decaying (2007bi-like) SLSNe are powered by the same mechanism, and whether there are two distinct classes or a continuum of events, remains to be seen.

A related phenomenon is the ``pulsational pair-instability'' \citep{woo2007} in stars of 65--130
\M. In this case, the energy released by explosive burning, following
pair-production, is less than the binding energy of the star. Many
solar masses of material may be ejected before the star resumes stable
burning, and the instability may be
encountered several times before a normal core-collapse SN terminates
its life. This is a promising means of producing circumstellar shells
(H-rich or -poor) in interaction models of SLSNe. No definitive objects of this type have been identified, but \citet{ami2014} have presented SN 2010mb, an energetic SN Ic (though not technically super-luminous) with an extremely extended light curve and narrow oxygen emission lines, and their analysis gave strong evidence for a SN interacting with hydrogen-free circumstellar material, matching predictions of pulsational-PISN models.

\begin{figure}
\includegraphics[width=8cm,angle=0]{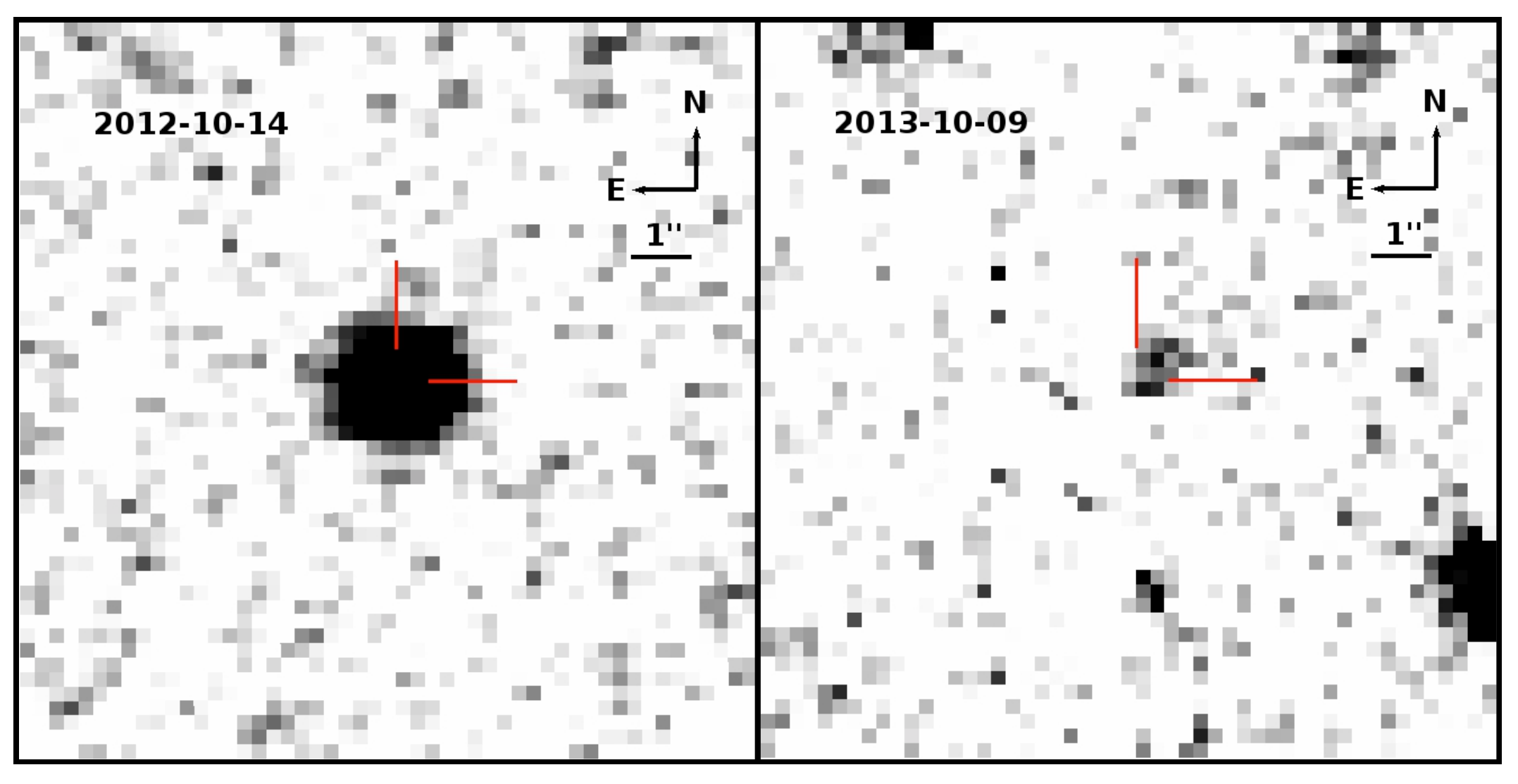}
\caption{NTT+EFOSC2 $V$-band images of LSQ12dlf (200s; {\it left}) and a faint, extended source at the SN location, likely to be the host galaxy (18$\times$200s; {\it right}).}\label{dlf_host}
\end{figure}

\begin{figure}
\includegraphics[width=8cm,angle=0]{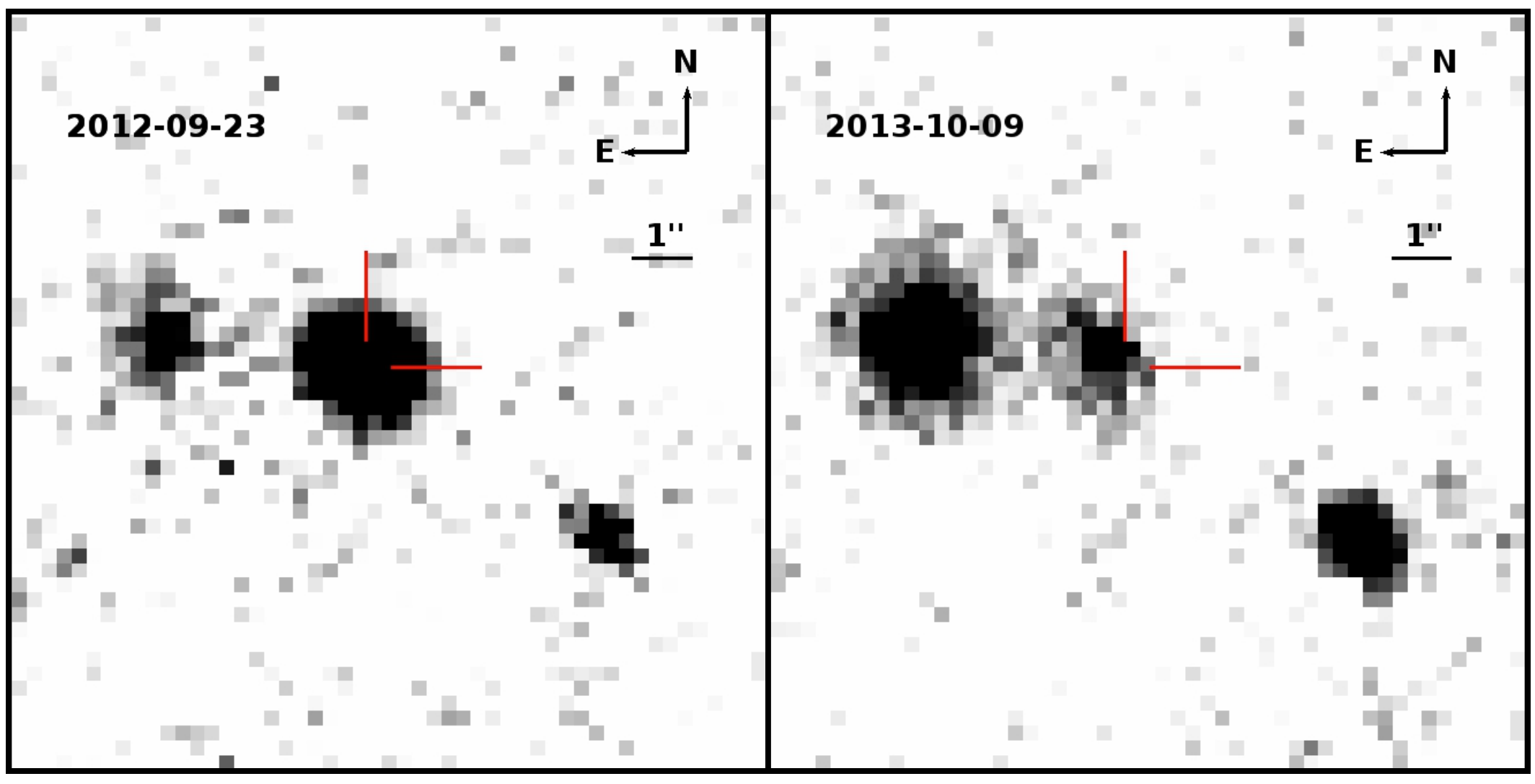}
\caption{NTT+EFOSC2 $R$-band images of SSS120810 (60s; {\it left}) and its host galaxy (18$\times$200s; {\it right}). The SN is offset $\sim0.\arcsec5$ from the centre of the host.}\label{sss_host}
\end{figure}

The Public ESO Spectroscopic Survey of Transient Objects (PESSTO; \citet{sma2014}; Smartt et
al., in prep) aims to classify and follow up hundreds of young and
unusual SNe, including those of the super-luminous variety, using
primarily the European Southern Observatory (ESO) 3.58m
New Technology Telescope (NTT) and EFOSC2 spectrograph \citep{buz1984}. These spectra are publicly available on WISeREP\footnote{http://www.weizmann.ac.il/astrophysics/wiserep/}  \citep[Weizmann Interactive Supernova data REPository;][]{yar2012}. PESSTO
provides a unique opportunity in the study of these rare objects,
since the survey strategy is naturally geared towards classifying
young objects and guaranteeing follow-up spectra with good
time-sampling. In this
paper, we report observations and modelling of three
SLSNe Ic discovered during the first year of PESSTO: LSQ12dlf,
SSS120810:231802-560926, and SN 2013dg, and also apply our models to PTF12dam \citep{nic2013}, SN 2011ke \citep{ins2013}, and the
SLSN II, CSS121015:004244+132827 \citep{ben2014}. In section
\ref{disc}, we describe the discovery and classification of each SN. 
Section \ref{spec} presents and discusses their spectra,
while section \ref{phot} does the same for the light curves. We
have developed a suite of light curve fitting tools, which we outline in section \ref{models}; these models are then applied in section \ref{fits}. We
summarize our findings in section \ref{conc}.

\section{Discovery and classification}\label{disc}

\subsection{LSQ12dlf}

LSQ12dlf was identified as a hostless transient by the La Silla QUEST
Variability Survey \citep[LSQ;][]{balt2013},
using the ESO 1.0m Schmidt Telescope, on 2012
July 10.4 UT, at RA=01:50:29.8, Dec=-21:48:45.4 (all coordinates in this paper are given in J2000.0). A spectrum obtained
by PESSTO with NTT+EFOSC2, on 2012 Aug 08.3 UT, showed it to be a
SLSN Ic about 10 days after peak luminosity. Comparison with SN
2010gx, and the other members of the PESSTO SLSN sample, indicated a
redshift $z\approx0.25$ \citep{12dlf_atel}. No host galaxy emission or
absorption lines are visible, even in a higher-resolution follow-up spectrum obtained with
 the Very Large Telescope (VLT)+X-Shooter \citep{ver2011}.  To determine the redshift, we 
cross-correlated the X-shooter spectrum (which we found was at an epoch of 
+36d after maximum, see Section\,\ref{spec-evol}) with a spectrum of SN 2010gx 
at +29d \citep{pas2010}. 
We found a minimum relative shift in the cross-correlation function for a redshift of $z=0.255\pm0.005$. 
Deep EFOSC2 imaging on 2013 Oct 10.3 UT, $\sim$300 days
after peak in the SN rest frame, and further follow-up in Jan-Feb 2014,
showed a very faint host galaxy, with a magnitude $V\approx25$ (Figure \ref{dlf_host}).

\begin{figure*}
\includegraphics[width=16cm,angle=0]{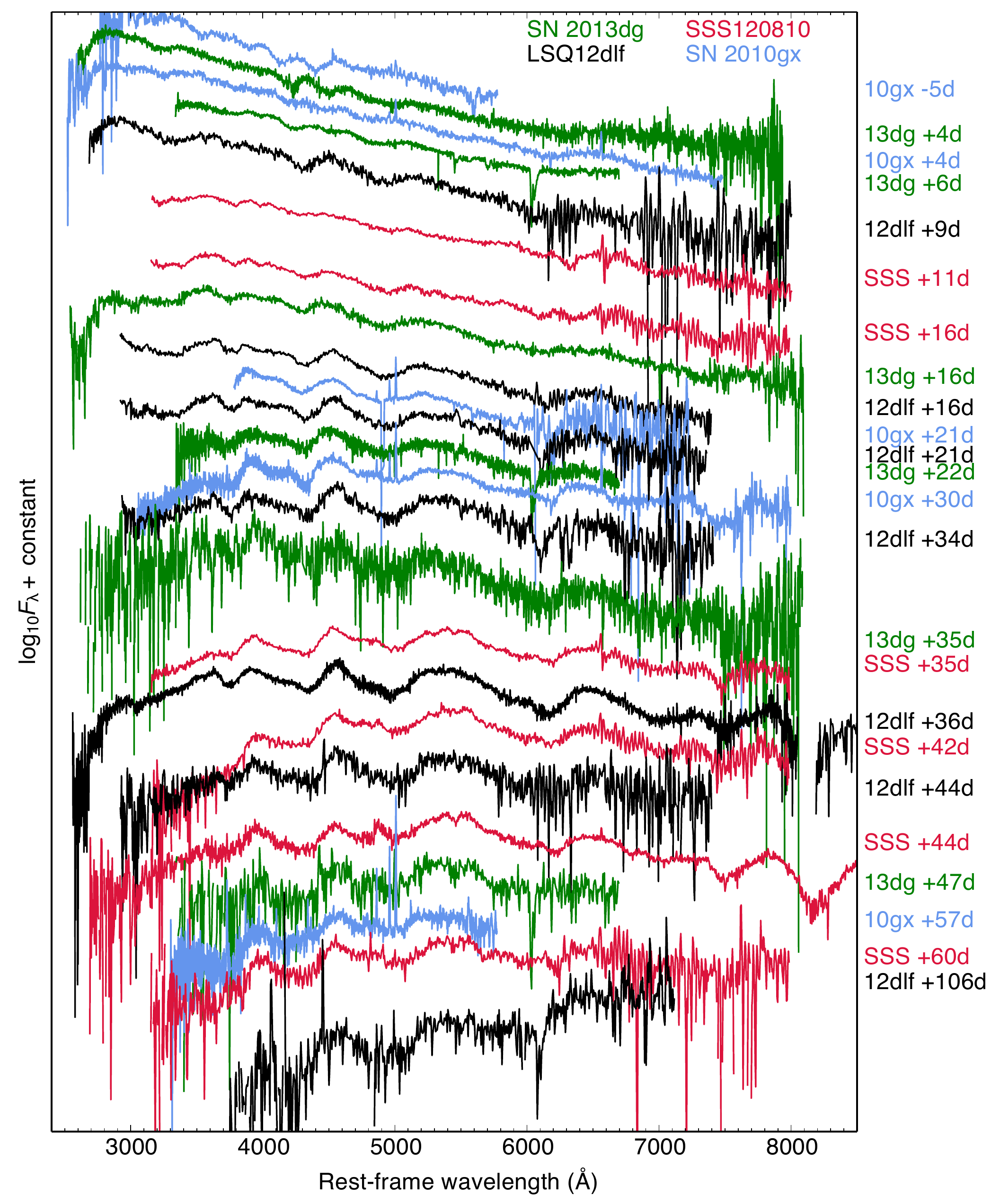}
\caption{The spectral evolution of our three PESSTO SLSNe, compared to
  SN 2010gx, a well-observed SLSN Ic. The four SNe show virtually
  identical spectral evolution, dominated initially by blue continua,
  and then by broad lines of Ca II, Mg II, Fe
  II and Si II (see Fig. \ref{xshoo}). Epochs (RHS) given in days from light curve peak, in
  the rest-frames of the SNe.}\label{spec_fig}
\end{figure*}

\begin{figure*}
\includegraphics[width=18cm,angle=0]{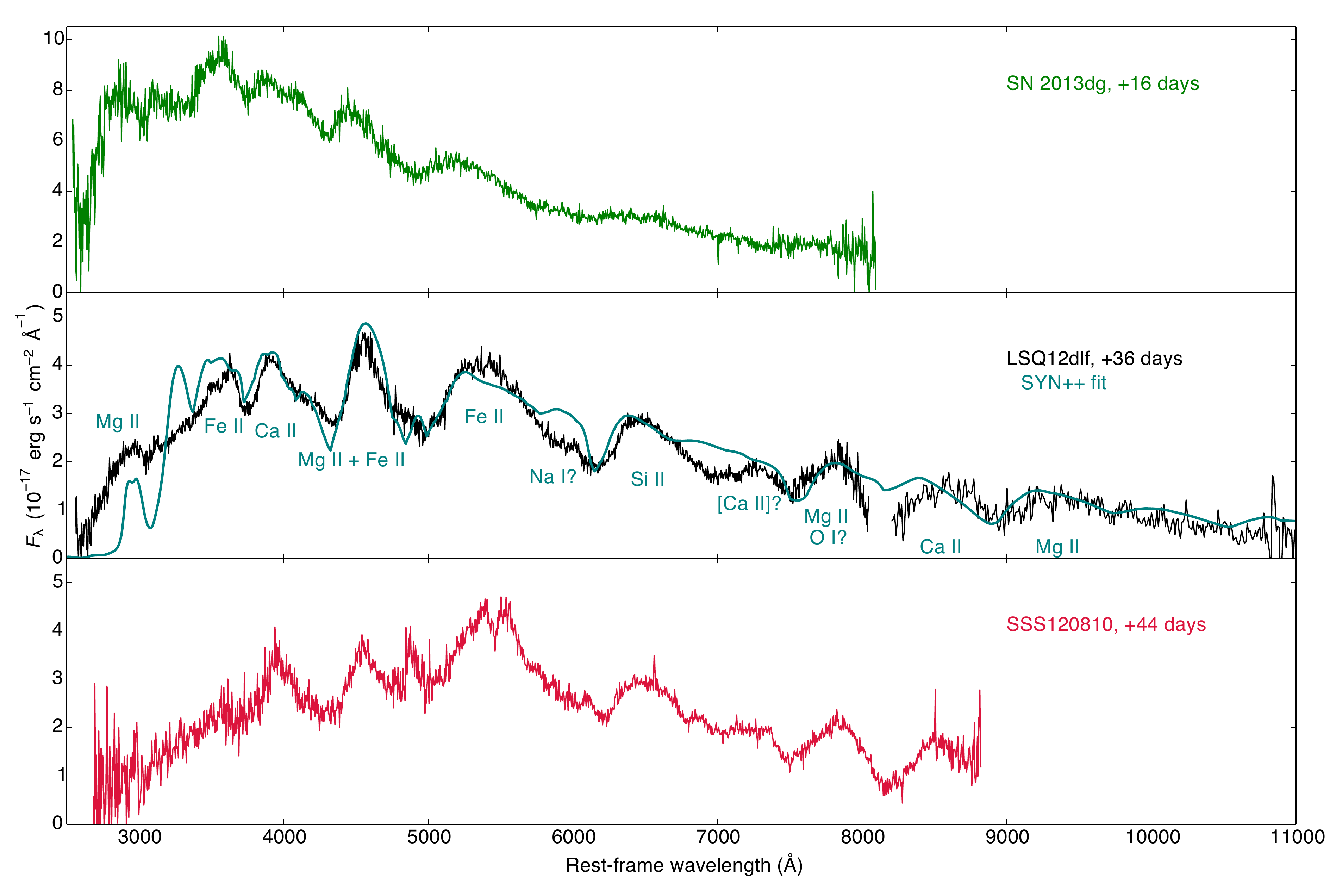}
\caption{VLT+X-shooter spectra of our 3 SLSNe, at 2--6 weeks after maximum
  light (rest-frame). The data have been binned by 10 pixels to improve
  the signal-to-noise. Also shown is a synthetic spectrum, generated using SYN++ \citep{tho2011}, used to identify the main line-forming ions. The best-fitting model spectrum has a photospheric velocity of $11000\,\rm{km}\,{s}^{-1}$, and temperature of 7000 K. We see that the spectra are dominated by singly-ionised intermediate-mass metals and iron.}\label{xshoo}
\end{figure*}

\subsection{SSS120810}
 
SSS120810:231802-560926 (hereafter: SSS120810) was discovered by the Siding Spring Survey (SSS), a division of the Catalina Real-Time Transient Survey \citep{dra2009}
with the 0.5m Uppsala Schmidt Telescope, on 2012 Aug 11.2. No host was present in SSS reference images at the location of the SN (RA=23:18:01.8, Dec=-56:09:25.6). PESSTO classified it on 2012 Aug 17--18 as a SLSN Ic, again roughly 10 days after maximum light \citep{12dlf_atel}. The redshift was initially estimated as $z\approx0.14-0.16$, from comparisons with other SLSNe Ic.  A spectrum taken at +44d after peak with the VLT+X-shooter (see Section \ref{spec-evol}), showed a distinct, narrow emission line at 7587.5 \AA, with a full-width-half-maximum of FWHM=9.4 \AA. The line is resolved and is almost certainly  \Ha~at $z=0.156$. Unfortunately, this is right in the telluric A band, which compromises a definitive measurement of the flux and width. Assuming this redshift, the X-Shooter spectrum also shows weak and narrow lines at wavelengths corresponding to two other common host galaxy emission lines : [O II] $\lambda$ 3727 and  [O III] $\lambda$ 5007. This gives confidence that the strongest narrow emission line is indeed
 \Ha~at $z=0.156$ and we adopt this redshift for the supernova. 
 Deep $BVRI$ imaging with EFOSC2 on 2012 Oct 10.1 and 25.0, $\sim$380 rest-frame days after peak, revealed a clear host galaxy, which is likely the source of the narrow emission lines. The SN is offset from the centre of this galaxy by $0.\arcsec51\pm$0.04 (Fig. \ref{sss_host}).

\subsection{SN 2013dg}

SN 2013dg was detected by the Mount Lemmon Survey (MLS) and Catalina Sky
Survey (CSS), both of which are part of CRTS \citep{dra2009}
. MLS initially discovered
the transient, MLS130517:131841-070443, on 2013 May 17.7 UT with the
1.5m Mt. Lemmon Telescope, while CSS independently found it with
the 0.68m CSS Schmidt Telescope on
May 30.7 UT, giving the alternative designation
CSS130530:131841-070443. The exact coordinates are RA=13:18:41.38 and Dec=-07:04:43.1. PESSTO identified MLS130517 as an interesting
target, but could not take an EFOSC2 spectrum at this time, as the
survey takes a break from May--July when the Galactic centre
is over La Silla. We instead classified this object using the William
Herschel Telescope (WHT) and ISIS spectrograph on 2013 Jun 11.0. The spectrum was dominated by a
blue continuum, and resembled SLSNe Ic a few days after maximum light \citep{13dg_atel}.
The WHT spectrum has two features that were identified as possible Mg II absorption, from either 
the host galaxy of SN 2013dg or intervening material, at a redshift of $z = 0.192$  \citep{13dg_atel}.
These are not visible in the X-shooter spectrum, but the signal-to-noise of the data at 3300 \AA~precludes
a useful quantitative check. The features are at the correct separation if they were the Mg II $\lambda\lambda$
2795.528, 2802.704 doublet, but the data are noisy, the lines are close to edge of the CCD and they can't be 
confirmed as real. Nevertheless, this redshift is ruled out by the broad supernova features in the spectrum. 
 We  cross-correlated the X-shooter spectrum (which we found was at an epoch of 
+16d after maximum, see Section\,\ref{spec-evol}) with a spectrum of SN2010gx 
at +11d \citep{pas2010}. We found a minimum relative shift in the cross-correlation function when we set $z=0.265\pm0.005$.  Hence we suspect the possible absorption is either not real or 
is foreground and we and adopt a redshift of  $z=0.265\pm0.005$ for the supernova. 
No host galaxy emission lines are visible in any of our spectra and the host is not detected in deep imaging
taken 250 days after peak (in Feburary 2014) down to $r>25.6$. 

\section{Spectroscopy}\label{spec}

\subsection{Data aquisition and reduction}

 The majority of our spectroscopy was carried out within PESSTO, using
 NTT+EFOSC2. The data were reduced using our custom PESSTO pipeline
 (developed in {\sc python} by S. Valenti), which calls standard {\sc iraf}\footnote{Image Reduction and Analysis Facility (IRAF) is distributed by the National Optical Astronomy Observatory, which is operated by the Association of Universities for Research in Astronomy, Inc., under cooperative agreement with the National Science Foundation.} tasks
 through {\sc pyraf}, to 
 de-bias and flat-field the two-dimensional frames, and
 wavelength- and flux-calibrate the spectra using arc lamps and
 spectrophotometric standard stars, respectively. The spectra are
 cleaned of cosmic-ray contamination using {\sc lacosmic}
 \citep{dok2001} before the 1D spectrum is extracted. The pipeline also uses
 a model to subtract telluric features (see Smartt et al. in prep).

Each of our SLSNe were also observed with VLT+X-Shooter. These data were
reduced using the X-Shooter pipeline within ESO's {\sc Reflex}
package. X-Shooter routinely observes telluric standard stars for all targets, and these were used to remove telluric features from our spectra within {\sc iraf}. SN 2013dg was classified with WHT at the beginning of the PESSTO
off-season, and additional spectra were obtained with GMOS on the Gemini
South telescope \citep{hook2004}. These were processed using standard {\sc iraf} tasks in
{\sc ccdproc} and {\sc onedspec}; GMOS spectra were extracted using the {\sc gemini}
package, while the WHT spectrum was extracted with {\sc apall}. The details of all spectra can be found in Tables
\ref{lsq_spec}, \ref{sss_spec} and \ref{13dg_spec}.

\subsection{Spectral evolution}
\label{spec-evol}

The observed spectral evolution of our three objects is shown
in Figure \ref{spec_fig}. All spectra have been corrected for redshift
and Milky Way extinction, according to the recalibration of the infrared galactic dust maps
by \citet{schlaf2011} ($E(B-V)=0.013; 0.019; 0.047$, for LSQ12dlf, SSS120810 and SN 2013dg, respectively). We assume negligible internal
extinction, since narrow Na I D absorption features are always very weak or
absent. All phases are given in days, in the SN rest-frame, from the date of maximum luminosity.

Between $\sim$10--60 days after peak, we have excellent time-series coverage
of all three supernovae, which we compare to one of the most
thoroughly observed SLSNe Ic, SN 2010gx \citep{pas2010}. Our earliest spectrum is the WHT
classification of SN 2013dg, obtained at 4 days after
maximum light. At this phase, the spectra are
dominated by a blue continuum with a blackbody temperature $T\sim13000$K, with a
few weak absorption features between 4000--5000 \AA. These may be attributable to O II, which tends to dominate
this region of SLSNe Ic spectra before and around peak, as can
be seen in SN 2010gx. However, these lines seem to be at slightly redder
wavelengths in SN 2013dg. It is possible that we have observed SN
2013dg during the transition from the O II--dominated early spectrum
to the nearly-featureless spectrum (with broad, shallow iron lines) seen just
after peak in SN 2010gx. This interpretation is supported by our GMOS spectrum, 2 days later, which closely resembles the first post-maximum spectrum of SN 2010gx. It should be noted, however, that radiative transfer models by \citet{how2013} instead favoured C II/III and Fe III as the dominant species in the early (around maximum light) spectra of some SLSNe.

The spectral evolution over days 10--60 is remarkably consistent across these four
SNe. Our X-Shooter spectrum of LSQ12dlf, at 36 days after light curve maximum, is shown in Figure \ref{xshoo}, along with a synthetic SYN++ spectrum for line identification \citep{tho2011}. The spectrum is dominated by singly-ionised metals. The strongest lines in the optical are Ca II H\&K, Mg II $\lambda$ 4481 (blended with Fe II), a broad Fe II feature between $\sim4900-5500$ \AA, and the Si II
$\lambda$ 6350 doublet feature. Beyond a phase of $\sim$50 days, the peak of the feature around $\sim$4500 \AA~appears to move to slightly redder wavelength, suggesting Mg I] $\lambda$ 4571 emission is dominant. These are the same features
identified by \citet{ins2013} in a sample of SLSNe Ic at similar
redshifts.

The Fe II lines in the spectrum of LSQ12dlf are stronger and develop earlier than in
the other objects of the sample, and are already quite pronounced
less than 10 days after peak light. This is very similar to the
behaviour of PTF11rks, in the \citet{ins2013} sample. That object
also transitioned to resemble a normal SN Ic by
this epoch, compared to the 20--30 days required in most SLSNe
Ic. The authors suggested that this faster evolution could be
related to its lower luminosity, relative to the other members of
their sample. However, LSQ12dlf peaks at an absolute AB
magnitude of $r\sim -21.4$, which is quite typical for SLSNe Ic, and in
line with the rest of our sample, and in fact declines more slowly in
luminosity than the other PESSTO objects (see Section \ref{phot}).

The final spectrum of LSQ12dlf,
106 days after maximum light, was taken with GMOS. By this time, the
SN has cooled to only a few thousand kelvin. The spectrum at this
epoch seems to be dominated by a fairly red continuum (i.e. it has not
reached the nebular phase), as well as the same broad lines of
singly-ionized metals that have remained present throughout the
observed lifetime of the SN. This latest spectrum may also show a weak
Na I D absorption; however, given the low signal-to-noise, such an
identification is not firm. The reddening is also present in our $k$-corrected photometry, which suggests $B-V\approx0.04$ at 44 days after peak, and $B-V\approx0.58$ at $\sim$100 days (the epochs of our last two spectra).

In the last spectrum of SSS120810, at 60 days, we do detect a weak Na I D
line. The spectra of SSS120810 beyond 35 days all show some
barely-resolved structure in the iron blends between $\sim$4500--5500
\AA, matching that in SN 2011kf \citep{ins2013}. X-Shooter spectra of SSS120810 and LSQ12dlf extend the wavelength range into the near-infrared. We can see strong Mg II absorptions at around 7500 and 9000 \AA, the former of which is probably blended with O I $\lambda$ 7775, and a clear Ca II NIR triplet. Overall, the spectral evolution of our objects seems to be much in line with the general picture of SLSNe Ic that has been emerging over the last few years.

% phot fig here

\begin{figure}

\begin{center}
	\subfigure{
		\includegraphics[width=8cm,angle=0]{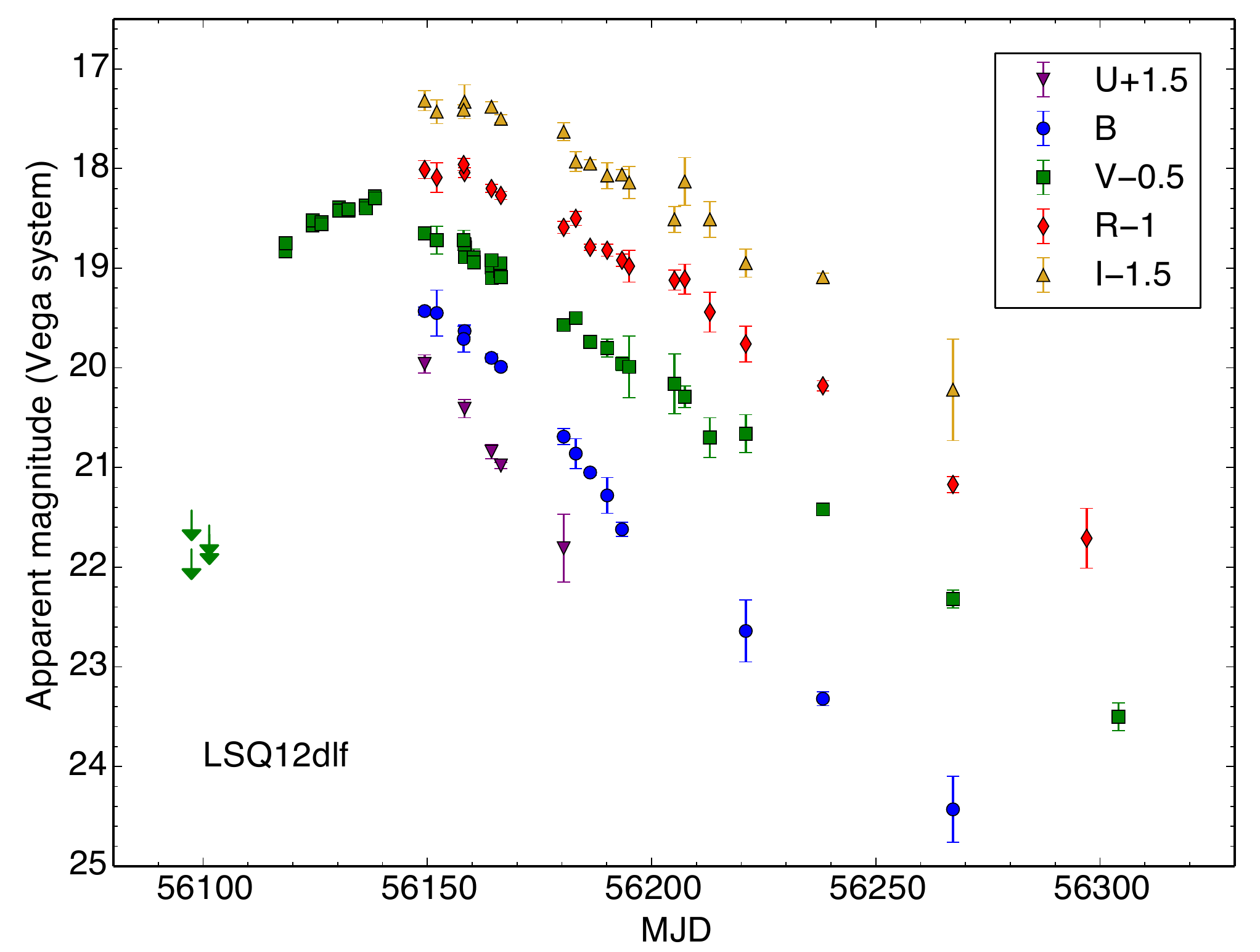}
		}\\
	\subfigure{
		\includegraphics[width=8cm,angle=0]{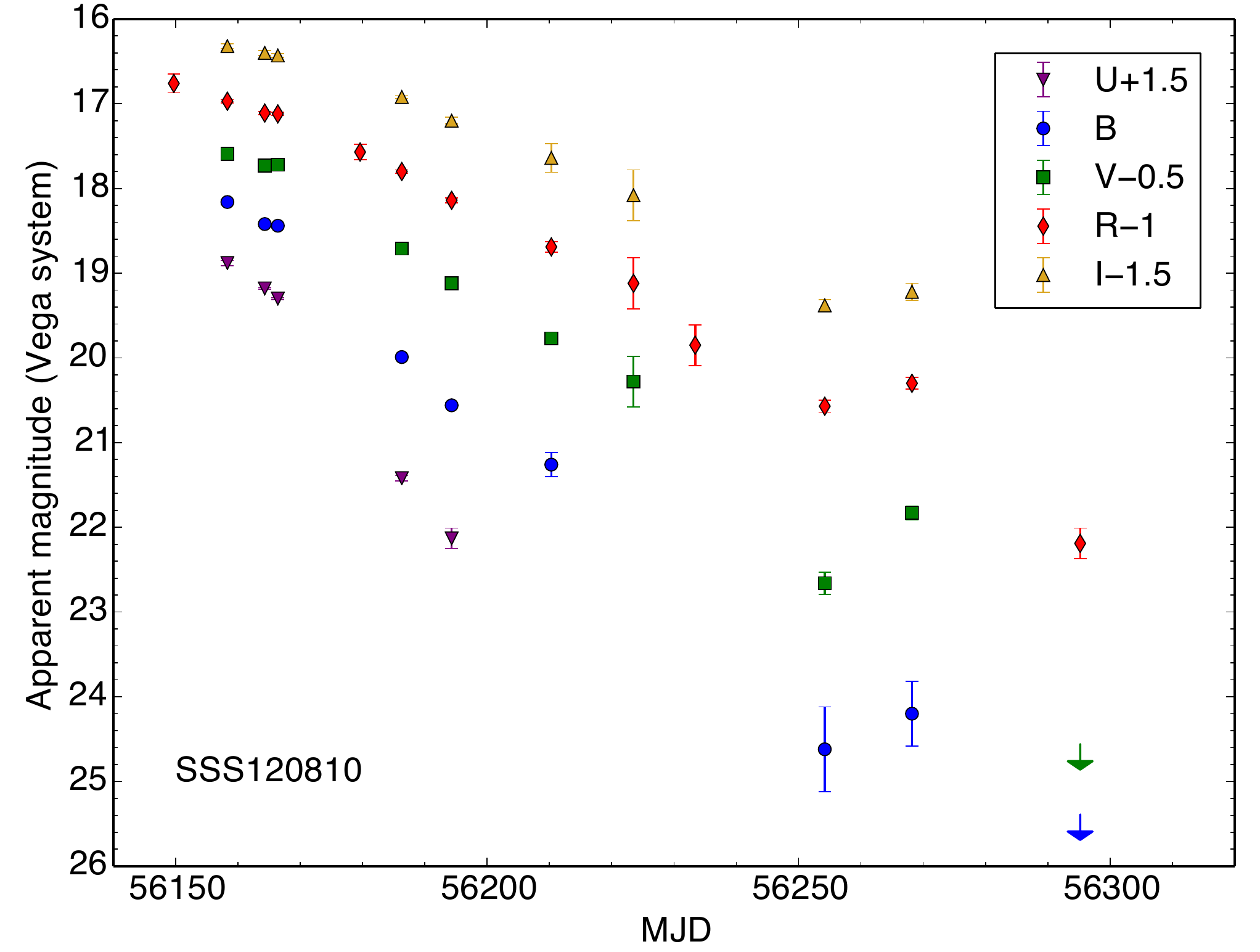}
		}\\
	\subfigure{
		\includegraphics[width=8cm,angle=0]{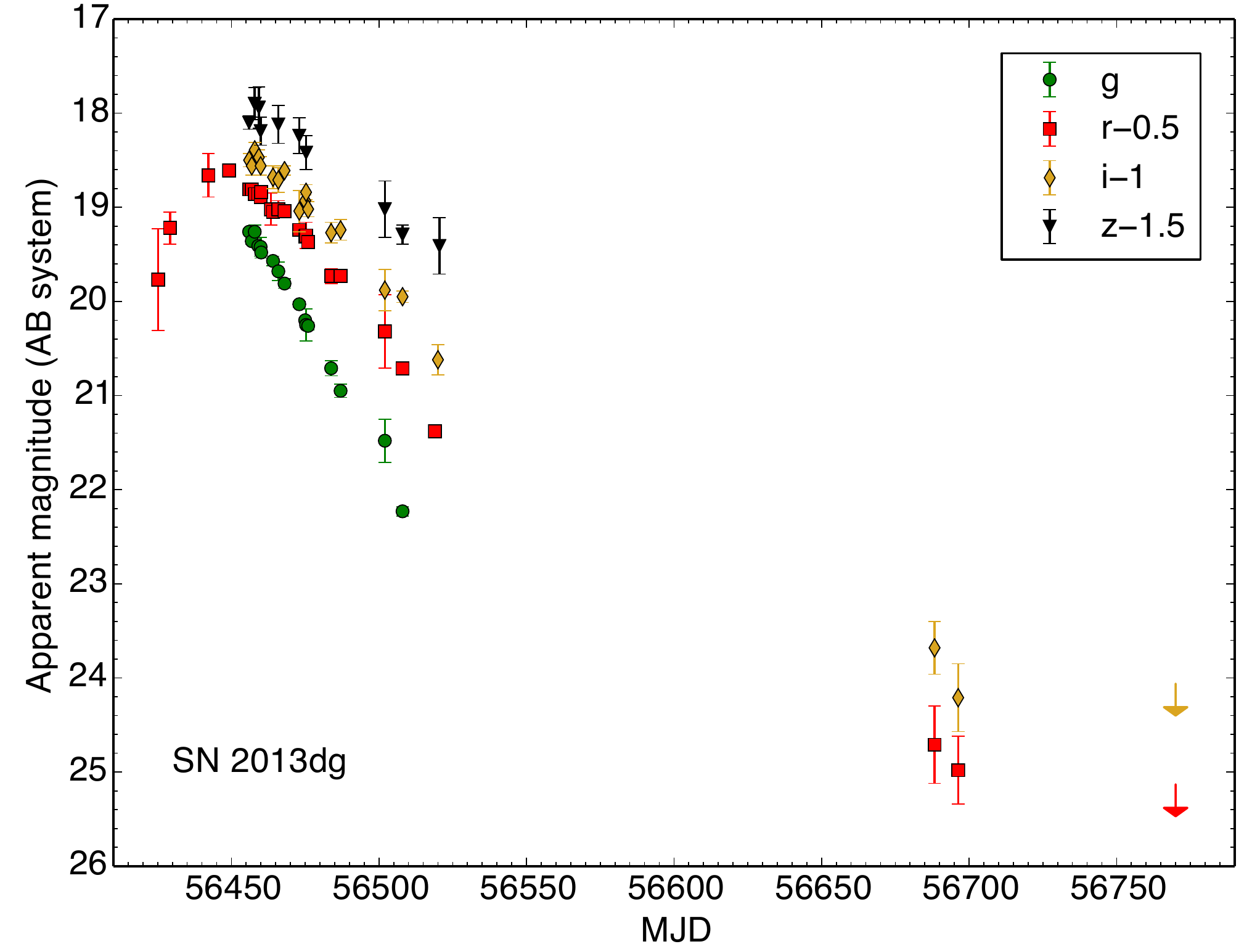}
		}\\
\end{center}
\caption{Multi-colour light curves of our SLSNe, in the observer frame}
\label{lcs}
\end{figure}

%\FloatBarrier

\section{Photometry}\label{phot}

\subsection{Data aquisition and reduction}

Imaging of our SLSNe came from a variety of sources. In addition to
NTT+EFOSC2, we collected data with the 2m Liverpool Telescope \citep{ste2004}, the Las Cumbres Observatory Global Telescope (LCOGT) 1m network \citep{bro2013}, and the 2m
Faulkes Telescopes (operated by LCOGT). Bias and flat field
corrections were applied using pipelines specific to each
instrument. SN magnitudes were measured by PSF-fitting photometry, while zero points were calculated using a local sequence of nearby stars (themselves calibrated to standard fields over several photometric nights).

Our light curves were supplemented with early data provided by LSQ, and public data from CRTS, which allow a determination of the rise times of these two SNe. Synthetic photometry on our
spectra showed that magnitudes calculated from LSQ images are almost
identical to those in the $V$ band, apart from a shift of $-0.02$ magnitudes to convert from LSQ AB mags to the more standard Vega system. For the public CRTS data, which are in the $R$-band, we averaged the (typically four) measured magnitudes from each night. We used the measured colour at peak, $r-R\approx0.2$, to convert to SDSS $r$ in the case of SN 2013dg. The EFOSC2 $i$ filter is closer to SDSS than to Johnson-Cousins; however for LSQ12dlf and SSS120810, we calibrate to Johnson-Cousins $I$ (Vega system) in keeping with the $UBVR$ photometry. For SN 2013dg, we used the SDSS-like $griz$ filters on EFOSC2, and calibrated to the AB magnitude system, in order to stay consistent with the Liverpool Telescope and LCOGT data obtained during the PESSTO off-season. The
magnitudes we measured for the three SNe are reported
in Tables \ref{lsq_phot_tab}, \ref{sss_phot_tab} and
\ref{dg_phot_tab}, where the final column in each table lists the data
source.

\subsection{Light curves}

Figure \ref{lcs} shows the
multi-colour photometric evolution of our three objects. The earliest observations
from LSQ and CRTS captured the rising phases of LSQ12dlf and SN
2013dg, respectively, while unfortunately the rise of
SSS120810 was missed. Judging from the similarity in the spectra of
SSS120810 taken on 24 Aug 2012 with that of SN 2013dg from 26 Jun 2013,
we estimate that SSS120810 peaked around MJD 56146 (7 Aug 2012), so the earliest
detection is likely just after maximum. All three objects exhibit a more rapid post-peak
decline in the bluer bands, which is typical for SNe of this kind, and
should be expected as they expand and cool.

The light curve of
SSS120810 shows some unusual behaviour at $\ga$100 days
after peak: a rebrightening, which is more pronounced in the
blue. Such a feature has not been witnessed in any previous SLSN. The
host galaxy of SSS120810 contributes significantly to the observed
brightness at the critical late
epochs (beyond $\sim$70d after peak), so we have subtracted deep EFOSC2 images of the host,
obtained in the second PESSTO season, after the SN had
faded. Subtractions were carried out using the code {\sc
  hotpants}\footnote{http://www.astro.washington.edu/users/becker/v2.0/hotpants.html}
\citep[based on algorithms developed by][]{ala1998}. The rebrightening remains significant even after template subtraction, and the measured zero points and sequence star magnitudes are consistent within $<$0.1 mag between these nights; we therefore conclude that this is real.

The host galaxy
of LSQ12dlf, by contrast, is barely detected in our late imaging, and
hence image subtraction need not be applied. SN 2013dg was in solar conjunction during September 2013--January 2014, but we picked it up again between the end of January  and April 2014. We detected a faint source in January and February 
2014 at +190d and +196d after peak, and this disappeared at +254d in similarly deep imaging. Hence it is likely that the source at $\sim$200 days is SN 2013dg, which has faded by 250 days, leaving no detection of the host galaxy. 
This means that image subtraction is not needed for any of the data points for SN 2013dg. 
The two late detections then suggest that there is a tail phase for SN 2013dg, as observed for several SLSNe Ic in the \citet{ins2013} sample.

\begin{figure*}
\includegraphics[width=15cm,angle=0]{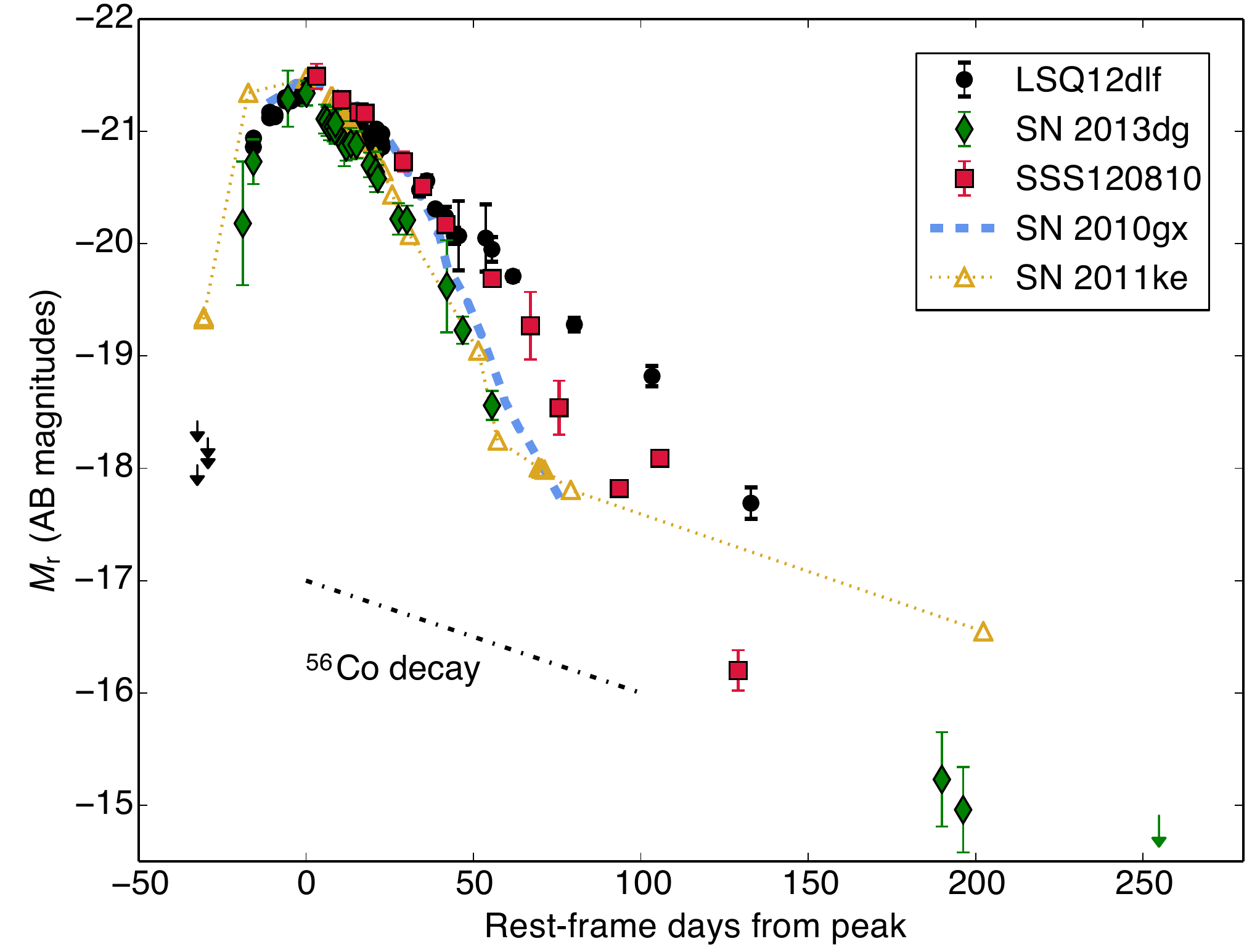}
\caption{Absolute r-band light curves. Phases have been corrected
  for time dilation, $K$-corrections (computed using synthetic photometry on our spectra, before and after correcting for cosmological expansion) have
  been applied to convert the effective filter to rest-frame $r$, and
  galactic extinction has been accounted for. The rising phases of
  LSQ12dlf and SN 2013dg are very similar, and last approximately
  25--35 days. However, our three objects show
  quite different declines after maximum light, though all fade more rapidly than fully-trapped \Co~decay. Also shown for
  comparison are SNe 2010gx and 2011ke \citep[typical SLSNe Ic,][]{ins2013}.}\label{lc_compare}
\end{figure*}

It is more instructive to directly compare the absolute light curves of
our sample. We assume a flat
$\Lambda$CDM cosmology, with H$_{0}=72\, \mathrm{km}\,
\mathrm{s}^{-1} \mathrm{Mpc}^{-1}$, $\Omega_{\Lambda}=0.73$ and
$\Omega_{\mathrm{M}}=0.27$. Time-dilation, galactic extinction corrections and
$K$-corrections have all been applied. The final $r$-band light curves
are shown in Figure \ref{lc_compare}, along with two other SLSNe Ic
\citep[2010gx and 2011ke,][]{pas2010,ins2013}. SN 2013dg is most similar
to the archetypal SLSNe Ic, including a likely flattening to a tail phase, albeit steeper than that seen 
in SN 2011ke at 50 days. LSQ12dlf rises with the same gradient as SN
2013dg -- the two SNe taking somewhere between 25--35 days to reach
peak -- but declines significantly more slowly. Moreover, LSQ12dlf shows
no sign of a break in the light curve slope even out to 130 days after
maximum. SSS120810 declines with a
slope intermediate between LSQ12dlf and SN 2013dg. No clear
radioactive or magnetar tail is seen; instead we see the rebrightening, which peaks at $\sim$ 100 days after maximum light.

\subsection{Bolometric light curves}\label{bol_text}

\begin{figure}
\includegraphics[width=8cm,angle=0]{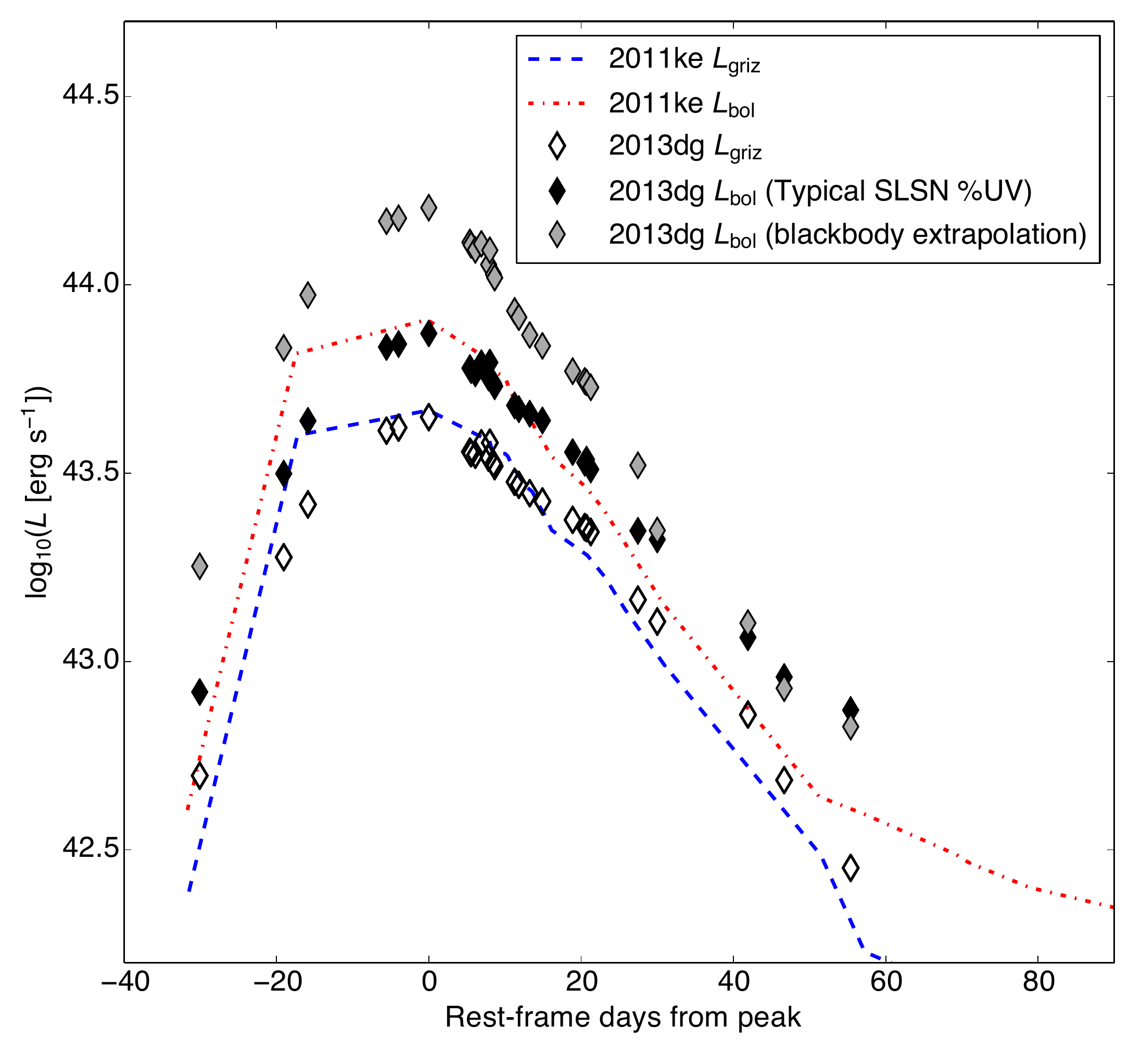}
\caption{The UV+NIR correction to the $griz$ pseudo-bolometric light curve of SN 2013dg, compared to the bolometric and $griz$ light curves of SN 2011ke \citep[from][]{ins2013}. We add the percentage UV flux, as a function of time, for a typical SLSN Ic \citep{ins2013}, to SN 2013dg, preserving the similarity in peak luminosity between these two SNe. We model the NIR flux by fitting a blackbody to the optical data. If we do the same for the UV flux, we overestimate the flux contribution from wavelengths shorter than rest-frame $g$-/$B$-band, as shown.}\label{uv_corr}
\end{figure}

In order to analyse our data using physical models, we have constructed a bolometric light curve for each of our SNe. First we de-reddened and $k$-corrected our photometry. We then integrated the corrected flux in these optical bands, and applied appropriate corrections for the missing ultra-violet and near-infrared data as follows: Initially, we tried applying corrections based on fitting a blackbody to the optical photometry, and integrating the flux between 1700 \AA~(approximately the blue edge of the Swift UVOT filters) and 25000 \AA. We compared the luminosity in the UV, optical and NIR regimes to the SLSNe studied by \citet{ins2013}, and found that we were likely significantly over-estimating the UV contribution, as is often the case with blackbody fits. In a real SN, UV absorption lines cause the flux to fall well below that of a blackbody at the optical colour temperature \citep{chom2011,lucy1987}. We therefore chose not to use the simple blackbody fit in the UV, and instead applied a typical percentage UV correction for SLSNe Ic, using Figure 7 of \citet{ins2013}. That work included both blackbody fits and real UV data, so should be slightly more reliable. The effect of this correction, and a comparison with SN 2011ke, is shown in our Figure \ref{uv_corr}. We did continue to use a blackbody estimate for the NIR contribution.

\begin{figure}
\includegraphics[width=8cm,angle=0]{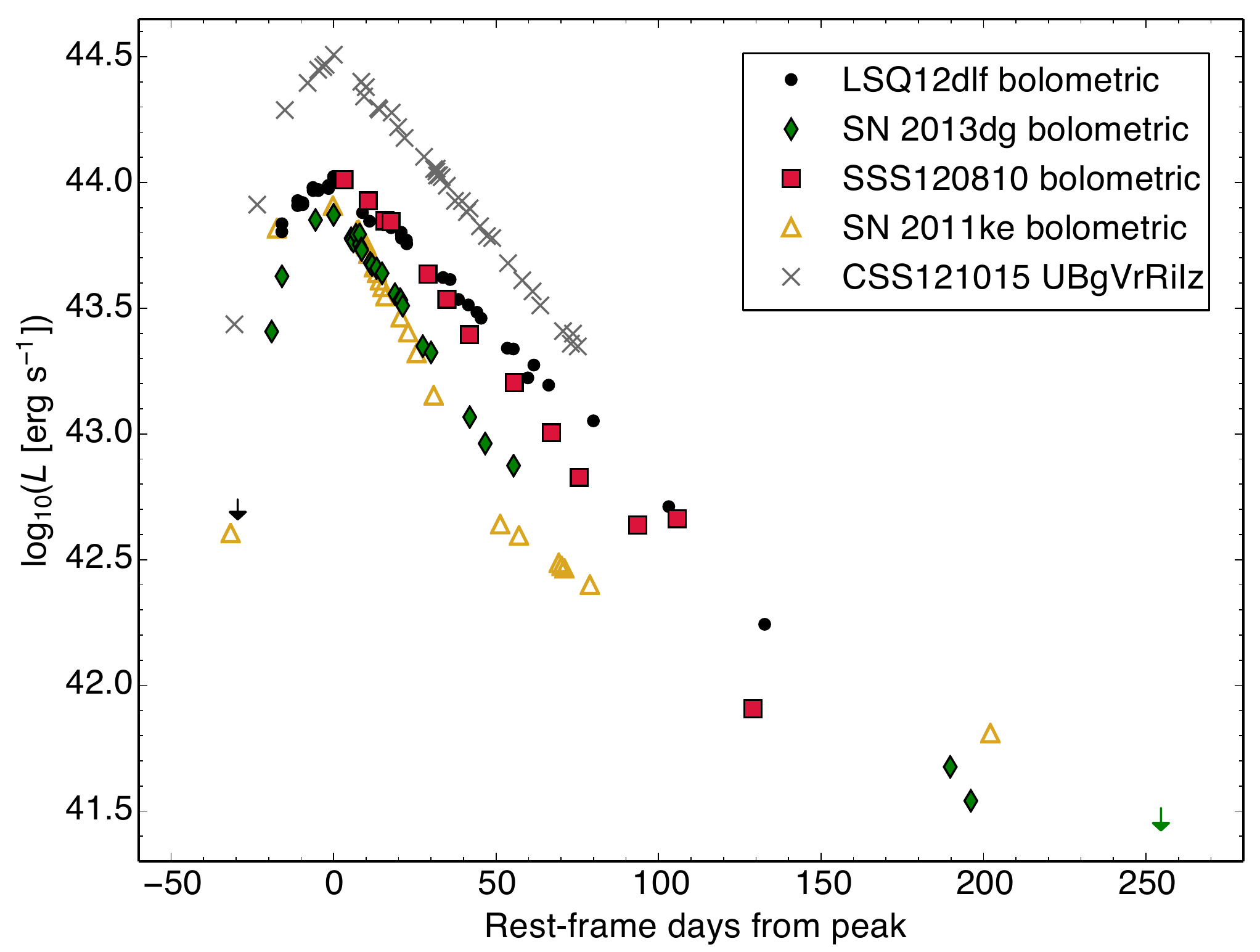}
\caption{Bolometric light curves of LSQ12dlf, SSS120810, SN 2013dg and SN 2011ke \citep[from][]{ins2013}, as well as CSS121015 \citep[a SLSN from PESSTO with spectral similarities to both SLSNe Ic and SLSNe II;][]{ben2014}.}\label{bol}
\end{figure}

Our bolometric light curves are shown in Figure \ref{bol}, along with SN 2011ke. Peak luminosities are in the range $0.7-1\times10^{44}\, \rm{erg}\,\rm{s}^{-1}$. As we saw in our single-filter comparison, the SNe have different declines from maximum. Of our three objects, only SN 2013dg shows evidence of a flattening in the decline rate, though not to the extent seen in SN 2011ke. The bolometric light curve of LSQ12dlf has a similar shape to that of another PESSTO SLSN, CSS121015 \citep{ben2014}, though it is significantly less luminous. CSS121015 showed evidence of circumstellar interaction, and the data were consistent with such a shocked-shell scenario, but it also exhibited similarities with SLSNe Ic, and its light curve was satisfactorily reproduced by one of our magnetar models. These two SLSNe show similar rise times and linear declines from peak magnitude. SSS120810 declines more rapidly than LSQ12dlf from a similar peak luminosity, though its linear decline is broken by the rebrightening at 100d. No published SLSN matches this behaviour. Possible mechanisms are proposed in section \ref{sss_disc}

\section{Light curve models for SLSNe}\label{models}

\subsection{Magnetar and radioactive models}

In previous works \citep{ins2013,nic2013, mcc2013}, we modelled SLSN Ic light curves using a semi-analytic code based on the diffusion solution of \citet{arn1982}, with radioactive nickel and magnetar radiation as power sources. Our magnetar model takes six parameters, of which the following four are free to vary: the ejected mass (\Mej), the magnetic field ($B$) and natal spin period ($P$) of the pulsar, and the explosion time ($t_{\rm shift}$). We fix the initial kinetic energy ($E_{\rm k}$) of the ejecta at $10^{51}$erg, and add to this the time-averaged energy input by the magnetar (minus that which is radiated away). The opacity ($\kappa$) is also fixed in our models, at $\kappa=0.2\,{\rm cm}^2\,{\rm g}^{-1}$. This choice of value is discussed in section \ref{csm_model}. The equations determining energy input and output are given in the appendix of \citet{ins2013}, and are based on the work of \citet{arn1982}, \citet{ost1971} and \citet{kas2010}. Good agreement has been found with the  detailed simulations of \citet{kas2010}. We fit these models to observed SLSN light curves by $\chi^2$ minimisation, after a coarse grid scan through parameter space has initialised the variables with sensible values.

The free parameters in our radioactive decay model are \Mej, the mass of radioactive \Ni~(\Mni), $E_{\rm k}$, and $t_{\rm shift}$. We again fix $\kappa$. In this case we omit the formal minimisation of $\chi^2$, since this almost invariably returns physically impossible fits, with \Mni \textgreater \Mej. Instead we iterate on a finer mass grid (with a resolution of 0.1 \M), for kinetic energies -- in units of $10^{51}$erg -- of 1, 3, 10 and 30 (if no satisfactory fit is obtained, we also try 100, i.e. in the case of CSS121015).

\subsection{Interaction model}\label{csm_model}

The models discussed above showed that the \Ni~decay chain struggles to reproduce SLSN Ic light curves, and that magnetar input can power the light curves observed so far. However, they did not allow us to comment quantitatively on the validity of strong interaction with circumstellar material (CSM) as an alternative power source. 
Mass-loss is a ubiquitous part of stellar evolution, especially in massive stars, and this mass-loss can build up a shell of gas around the star, which the SN ejecta then collide with. Developing a synthetic light curve tool based on ejecta-CSM interaction is therefore an important step in discriminating between the three main models for SLSNe (magnetars, radioactivity and interaction).

We have developed such a code by
implementing the formulae detailed in \citet{cha2012}. 
Their derivation assumes a stationary photosphere inside the CSM shell -- this is justified by the slow expansion velocity of the CSM relative to typical velocities in SNe. Energy is input efficiently by self-similar forward and reverse shocks (i.e. all of the kinetic energy of the shocks converts to radiation) generated at the ejecta-CSM interface \citep{che1982,che1994}, as well as by radioactive decay of \Ni~and \Co~deep in the ejecta. In this approximation, the time-dependence of the energy input from the shocks depends only on the density profiles of the interacting media (the ejecta and the CSM). The shock luminosity originates at the ejecta-CSM interface at all times in this model (to make the problem analytically tractable), but two important timescales are found: heat input by the forward shock is terminated abruptly when it breaks out of the CSM, and the reverse shock stops depositing heat when it has swept up all of the ejecta, leading to some discontinuity in the gradient of the light curve at these two epochs.

While the treatment of the shocks is based on that of \citet{che1982} and \citet{che1994}, those works dealt with an optically thin stellar wind, where we normally see X-ray emission from the reverse shock front, and strong narrow lines from pre-shock gas excited by these energetic photons. This is not the case for an optically thick CSM, which is necessary to explain the SLSNe \citep{smi2008}. In this regime, the diffusion of energy out of the shell is important, and we follow \citet{cha2012} in using the formalism of \citet{arn1980,arn1982}, in the special case of zero expansion velocity (our magnetar and nickel models use the same result with homologous expansion). Energy deposited by the shocks diffuses out of the region where the CSM is optically thick, whereas energy from radioactive decays must diffuse out of the combined mass of the ejecta and the optically thick CSM. Thus two different diffusion times are calculated. After shock heating ends, the solution for the light curve is governed simply by radiative diffusion from the opaque shell, unless there is significant heating from radioactivity.

We fit the observed light curves by $\chi^{2}$ minimisation, using mainly the same free parameters as those of \citet{cha2012}. The output luminosity is a function of ejected mass (\Mej), CSM mass (\Mcsm), nickel mass (\Mni), explosion time ($t_{\rm shift}$), ejecta kinetic energy ($E_{\rm k}$), interaction radius ($R_0$), CSM density ($\rho_{\rm CSM}$, as well as density scaling index, $s$), density scaling exponents for the SN core and envelope ($\delta$ and $n$, respectively), and the opacity ($\kappa$). Our code begins by scanning over a grid of points in this high-dimensional parameter space to look for the best approximate solution. This is then the starting point for a more rigorous $\chi^{2}$ minimisation, using the {\sc Python} module {\sc Scipy.Optimize.Fmin}.

To reduce the number of free parameters in our fits, we fix several variables at typical values. The most uncertain is perhaps the opacity. For hydrogen-free material, and when electron scattering is the dominant source of opacity, $\kappa$ is often taken to be $0.1\,{\rm cm}^2\,{\rm g}^{-1}$ \citep[see][and references therein]{ins2013}. For hydrogen-rich material, $\kappa=0.33\,{\rm cm}^2\,{\rm g}^{-1}$ may be more appropriate \citep{cha2012}. Since we do not know the composition of the CSM (but expect it to be H-poor, from the spectra we observe), we take an intermediate value, $\kappa=0.2\,{\rm cm}^2\,{\rm g}^{-1}$. For the sake of consistency, we use the same value in our magnetar and radioactive models. In non-interacting models, $\kappa$ enters into the light curve equation only in determining the diffusion timescale parameter: $\tau \propto \kappa^{1/2} M_{\rm ej}^{3/4} E_{\rm k}^{-1/4}$. As $\tau$ is what we really fit for, and $E_{\rm k}$ is either fixed or, in the case of the magnetar, determined from $B$ and $P$, for a given fit we have \Mej$\,\propto\,\kappa^{-2/3}$. Therefore, varying the opacity by a factor of 2 only changes the extracted mass estimate by a factor $\sim1.6$, which is not crucial to our analysis.

We fix $n=10$ and $\delta=0$ \citep{cha2012,che1994}, and test only $s=0$, corresponding to a uniform density shell produced by a massive outburst of the progenitor, and $s=2$, appropriate for a steady stellar wind prior to explosion. These were the cases studied by \citet{che1982}. There are still many remaining parameters, so \citet{cha2013} were not surprised to find a large degeneracy between them. \Mej~and $R_0$ were particularly weakly constrained, especially if both wind ($s=2$) and shell ($s=0$) models are considered. Because previous interaction-powered models of SLSNe have required very massive CSM, such that the mass loss rates would be extraordinarily high if this material were lost in a steady wind \citep[e.g. see the discussion in][]{ben2014}, for this paper we restrict our fits to uniform shells from large mass ejections ($s=0$). We also find that our fits are largely insensitive to the parameter $R_0$; it affects the light curve only insofar as it alters the radius of the photosphere (\Rph). We initially allowed $R_0$ to vary from $10^{12} - 10^{15}$cm, but the CSM shells we find in fits of SLSNe are all $\sim 10^{15}$cm thick, with the photosphere located close to the outer edge (as expected, since these shells are highly optically thick), so our `best-fit' $R_0$ is typically much smaller than \Rph, and can in fact be changed by factors of 10 or more with little to no effect on the other parameters of the light curve. We therefore fix $R_0$ at $10^{13}$cm ($\sim$150 ${\rm R}_\odot$) for simplicity. This leaves \Mej, \Mcsm, $E_{\rm k}$, $\rho_{\rm CSM}$, $t_{\rm shift}$, and (optionally) \Mni~as parameters to fit. We have 1--2 more free parameters in this model, compared to the magnetar model, so we expect that it will be easier to fit a wider range of light curve shapes.

The peak luminosity is most sensitive to $E_{\rm k}$ and \Mej, with more energetic or less massive explosions giving a brighter peak. The light curve timescales depend on $\rho_{\rm CSM}$, \Mej~and \Mcsm. Denser CSM results in a faster rise but slower decline, whereas more massive CSM tends to broaden the whole light curve, by increasing the diffusion time from the shell. More massive ejecta result in a slower rise, and can broaden the peak as it weakly increases the termination time for the forward shock while greatly increasing that for the reverse shock, but it has little effect on the final decline rate, as the shock energy is input at the base of the CSM shell (however, for significant nickel mass, \Mej~does affect the diffusion timescale). In most cases, the forward shock luminosity is greater than that of the reverse shock, such that the discontinuity at reverse shock termination is only visible in the light curve if this occurs after forward shock termination.

Of course, there are limitations to this analytical framework, many of which were also pointed out by \citet{cha2012}. The assumption of 100\% efficiency in converting kinetic energy to radiation is unrealistic for models with \Mcsm$<<$\Mej, and in this case we would also expect the photosphere to expand quickly, since the modest swept-up mass is insufficient to slow it down, in contrast to the fixed photosphere we use in the model. In our fits, we typically find  \Mcsm$\sim$\Mej$/2$, so these approximations are not bad for our purposes. In general, the dynamics of this situation are quite complex; however, comparisons between the analytic models and more realistic hydrodynamics simulations shown in \citet{cha2012} show that these simplified models are, at the very least, a useful guide to the regions of parameter space that can generate light curves of interest.

\section{Model fits to SLSN data}\label{fits}

%%%% Fit parameters

\begin{table*}
\caption{Light curve fit parameters}\label{fit_params}
\begin{tabular}{cccccccc}
\hline
\hline

{\bf Magnetar}\\
\hline
 & \Mej/\M & $B$/$10^{14}$G & $P$/ms & $\chi^2/{\rm d.o.f.}$ \\
\hline
CSS121015 & 5.5 & 2.1 & 2.0 & 10.39 \\
LSQ12dlf & 10.0 & 3.7 & 1.9 & 5.61 \\
SSS120810 & 12.5 & 3.9 & 1.2 & 10.63 \\
SN 2013dg & 5.4 & 7.1 & 2.5 & 1.01 \\
PTF12dam & 9.4 & 1.2 & 2.7 & 0.64 \\
SN 2011ke & 6.7 & 6.4 & 1.7 & 1.60 \\
\hline
\hline

{\bf \Ni~decay} \\
\hline
 & \Mej/\M & \Mni/\M  & $E_{\rm k}$/$10^{51}$erg & $\chi^2/{\rm d.o.f.}$ \\ 
\hline
CSS121015 & 20.3 & 20.2 & 100 & 329.14 \\
LSQ12dlf & 10.1 & 8.1 & 30 & 3.46 \\
SSS120810 & 7.2 & 6.6 & 30 & 10.45 \\
SN 2013dg & 6.6 & 5.5 & 30 & 0.37 \\
\hline
\hline

{\bf CSM interaction} \\
\hline
 & \Mej/\M & \Mcsm/\M & \Mni/\M & $E_{\rm k}$/$10^{51}$erg & log($\rho_{\rm CSM}$/${\rm g}\,{\rm cm}^{-3}$) & $\chi^2/{\rm d.o.f.}$ & \Rph/$10^{15}$cm (not fit) \\
\hline
CSS121015 & 6.7 & 4.9 & -- & 2.3 & -12.54 & 4.12 & 2.0 \\
LSQ12dlf & 7.6 & 3.4 & -- & 1.1 & -11.95 & 0.80 & 1.1 \\
SSS120810 & 15.8 & 2.3 & -- & 0.84 & -11.74 & 12.78 & 2.1 \\
SN 2013dg & 4.6 & 2.4 & -- & 1.2 & -12.22 & 0.38 & 0.6 \\
PTF12dam & 26.3 & 13.0 & -- & 1.9 & -12.06 & 0.45 & 2.6 \\
SN 2011ke & 0.8 & 0.1 & 0.3 & 0.1 & -15.07 & 13.95 & 2.4 \\
SN 2011ke ($t<50$ d) & 10.8 & 0.1 & -- & 0.07 & --9.86 & 0.26 & 2.6 \\
\hline

\end{tabular}

\end{table*}

In this section we apply our three simple light curve models to the data presented in section \ref{phot}. We consider magnetar-powered light curves using the same method and analytical treatment discussed and applied by us previously in \citet{ins2013}, \citet{nic2013} and \citet{mcc2013}. A similar model based on \Ni-powering, as implemented in those same works, is also presented. Those papers showed that magnetar models could reasonably reproduce the bolometric light curves as well as the temperature and velocity evolution. We now also investigate fits with the simple CSM interaction model presented in the previous section. We start by applying these model fits to the three new PESSTO SLSNe, and then put this in context by revisiting three well-studied SLSNe using the CSM alternative model: SN 2011ke \citep{ins2013}, PTF12dam \citep{nic2013} and CSS121015 \citep{ben2014}.

The best fitting light curve models for each SN are shown in Figures \ref{dlf_models}--\ref{12dam_11ke}, and the parameters of all fits are listed in Table \ref{fit_params}. Errors are approximately the same size as the circles. Triangles represent upper limits. In practice, we find that no \Ni~is required for the interaction-powered fits to any of our SNe. We have also measured velocities (by fitting Gaussian profiles to spectral lines; errors are the scatter in multiple fits) and temperatures (from the automated blackbody fits to our photometry, used to estimate UV and NIR corrections in section \ref{bol_text}). These are compared to the predictions of our light curve models. Magnetar and nickel models give us photospheric velocities and temperatures, as described in \citet{ins2013}; hence these curves end when the SN no longer has a well-defined photosphere (i.e. its atmosphere has become optically thin). The temperature is estimated in our interaction model simply by assuming that the output luminosity is blackbody emission from the photosphere, and therefore using $L=4\pi R_{\rm phot}^2 \sigma T^4$. As the location of the photosphere, \Rph, is fixed in this model, we simply have $T \propto L^{1/4}$.

\subsection{LSQ12dlf}

\begin{figure}
\begin{center}
	\subfigure{
		\includegraphics[width=8cm,angle=0]{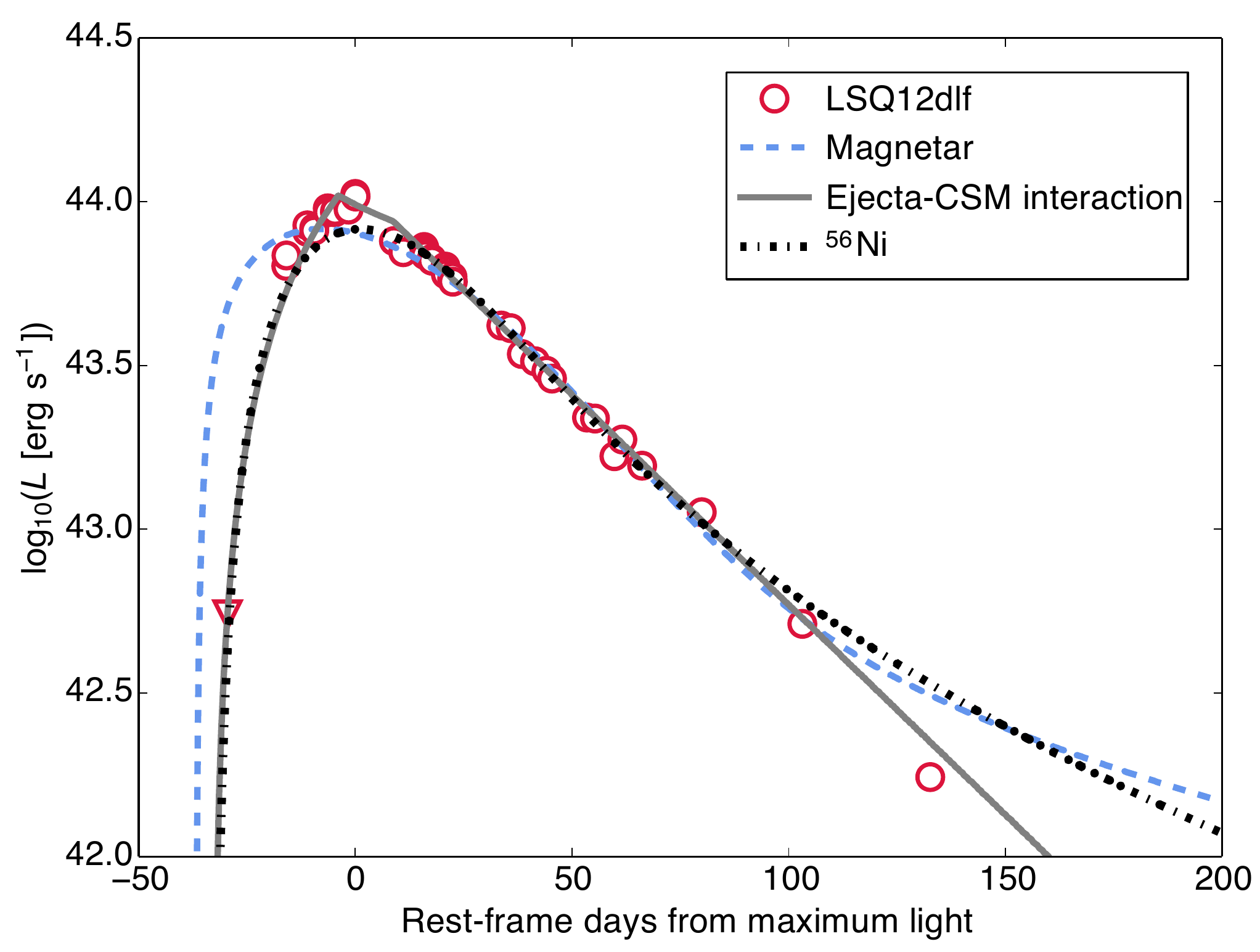}
		}\\
	\subfigure{
		\includegraphics[width=8cm,angle=0]{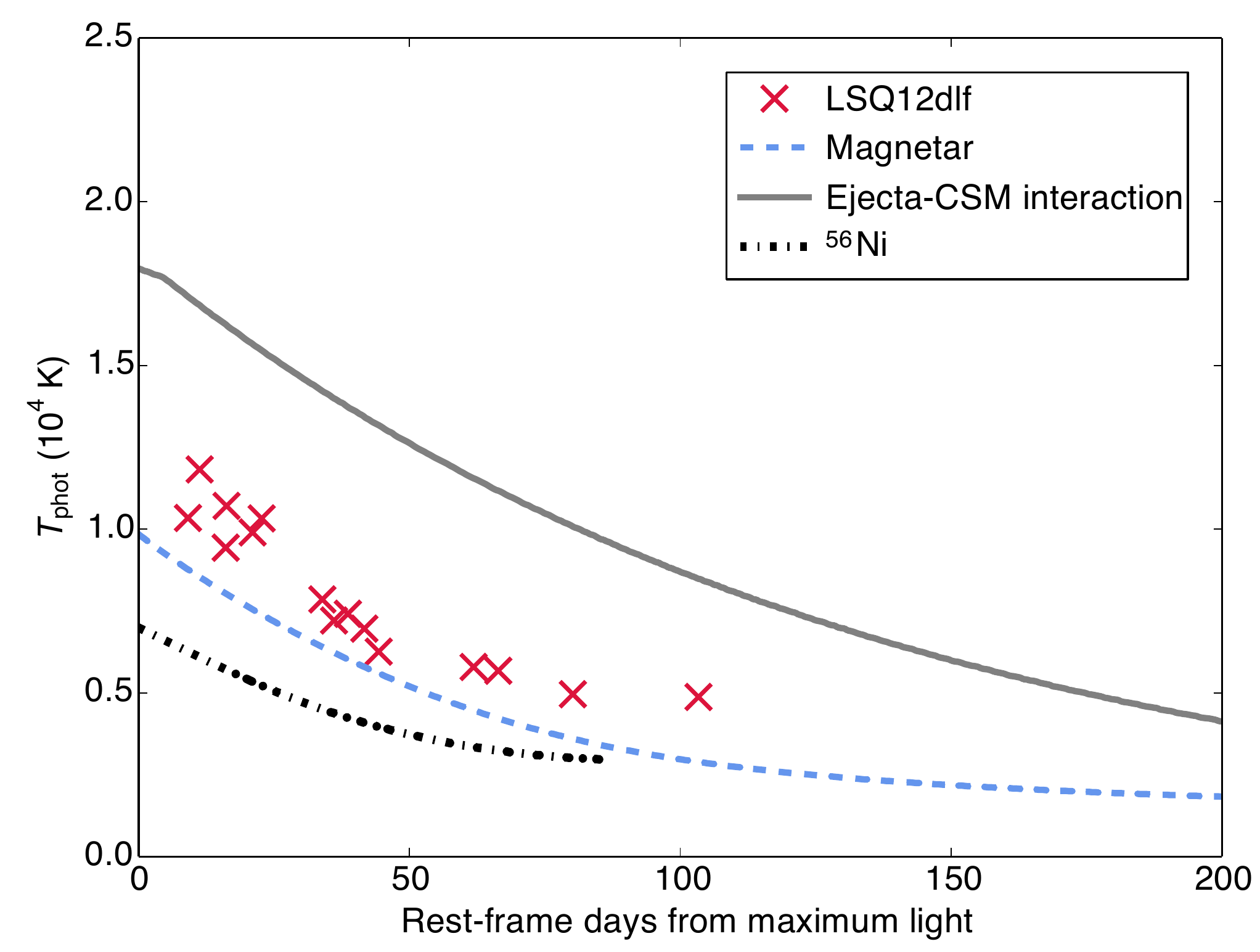}
		}\\
	\subfigure{
		\includegraphics[width=8cm,angle=0]{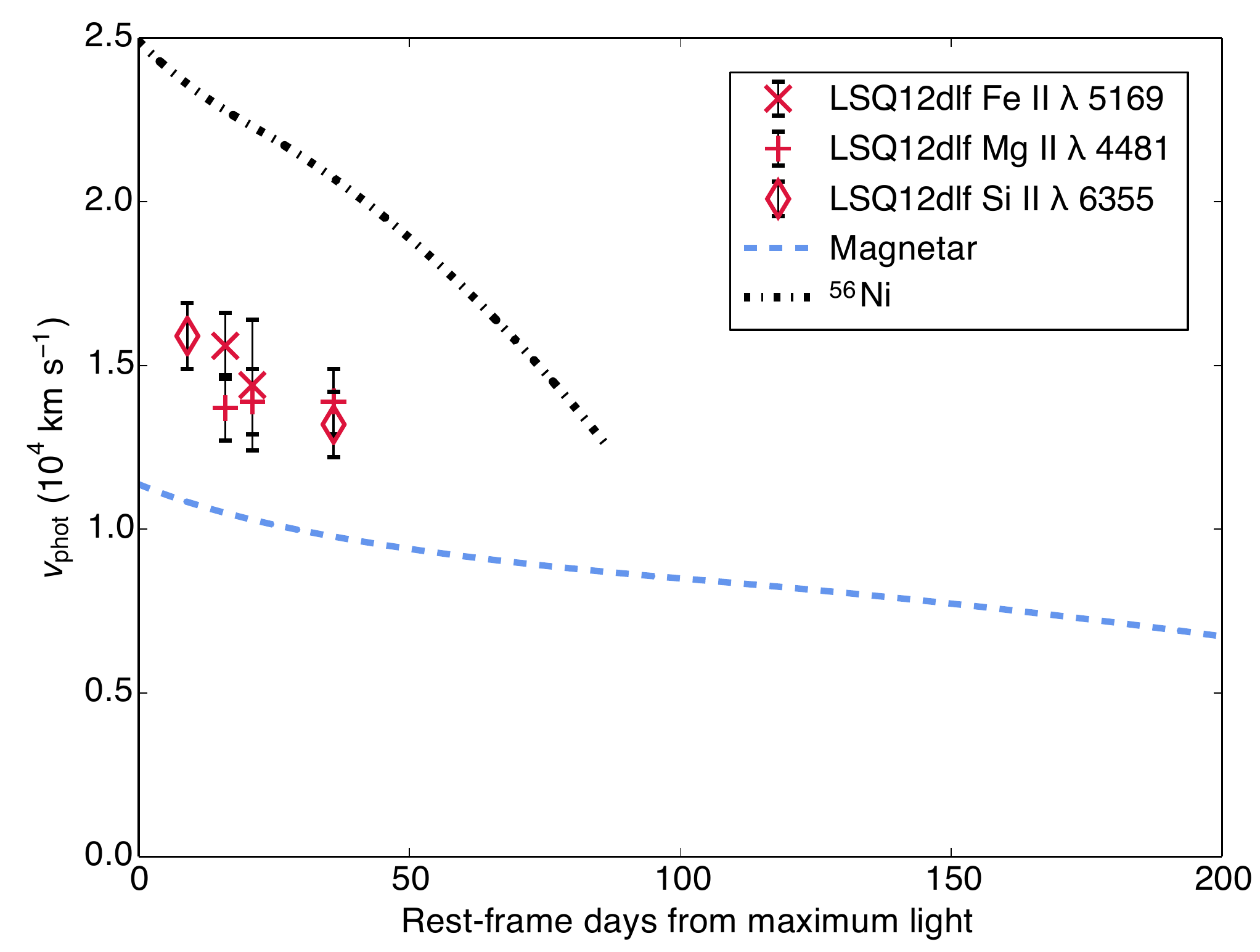}
		}\\
\end{center}
\caption{Magnetar-, interaction- and \Ni-powered models of LSQ12dlf. Parameters are listed in Table \ref{fit_params}. Temperatures were estimated by fitting blackbody curves to multi-colour photometry; velocities were measured from absorption minima.}
\label{dlf_models}
\end{figure}

In contrast to the SLSNe studied by \citet{ins2013}, LSQ12dlf (Fig. \ref{dlf_models}) is difficult to fit with a magnetar model. It has a noticeably broader light curve than all the other low-$z$ SLSNe Ic. The decline in magnitude is linear for $\sim$130 days, showing no sign of a $t^{-2}$ tail (or indeed a \Co~tail). Although the magnetar model fits the majority of the light curve well, the fit is poor at late times, where it over-predicts the flux, and early times, as fitting the slow decline results in a broader peak and, more importantly, an earlier explosion date than our limit at -30 days suggests. The peak is not so problematic, since these luminosities are estimated from single-filter LSQ photometry, and are therefore subject to significant uncertainty. The discrepancy between the magnetar model and the data at 130 days could be attributable to time-dependence of the magnetar energy-trapping. Most of the magnetar power is expected to be released in the form of X-rays/$\gamma$-rays and/or high-energy particle pairs; if the ejecta become optically thin to X-rays, for example, as the SN expands, the luminosity emitted as reprocessed optical radiation may drop below the predictions of our fully-trapped model. Therefore we cannot exclude the magnetar based on the late data point.

The interaction and radioactive models give a better fit to the early part of the light curve, though we exclude the latter because of the requirement for 80\% nickel ejecta. The interaction model is for $\sim8$ \M~of ejecta and $\sim3$ \M~of CSM. This gives a satisfactory fit to the whole light curve.

The magnetar model best matches the temperature evolution, however, all three models get the approximate shape correct. The nickel model greatly over-predicts the SN velocity, because of the large explosion energy ($30\times10^{51}$erg) needed to fit the light curve time scales. This is true for all of our SNe.

\subsection{SSS120810}\label{sss_disc}

\begin{figure}
\begin{center}
	\subfigure{
		\includegraphics[width=8cm,angle=0]{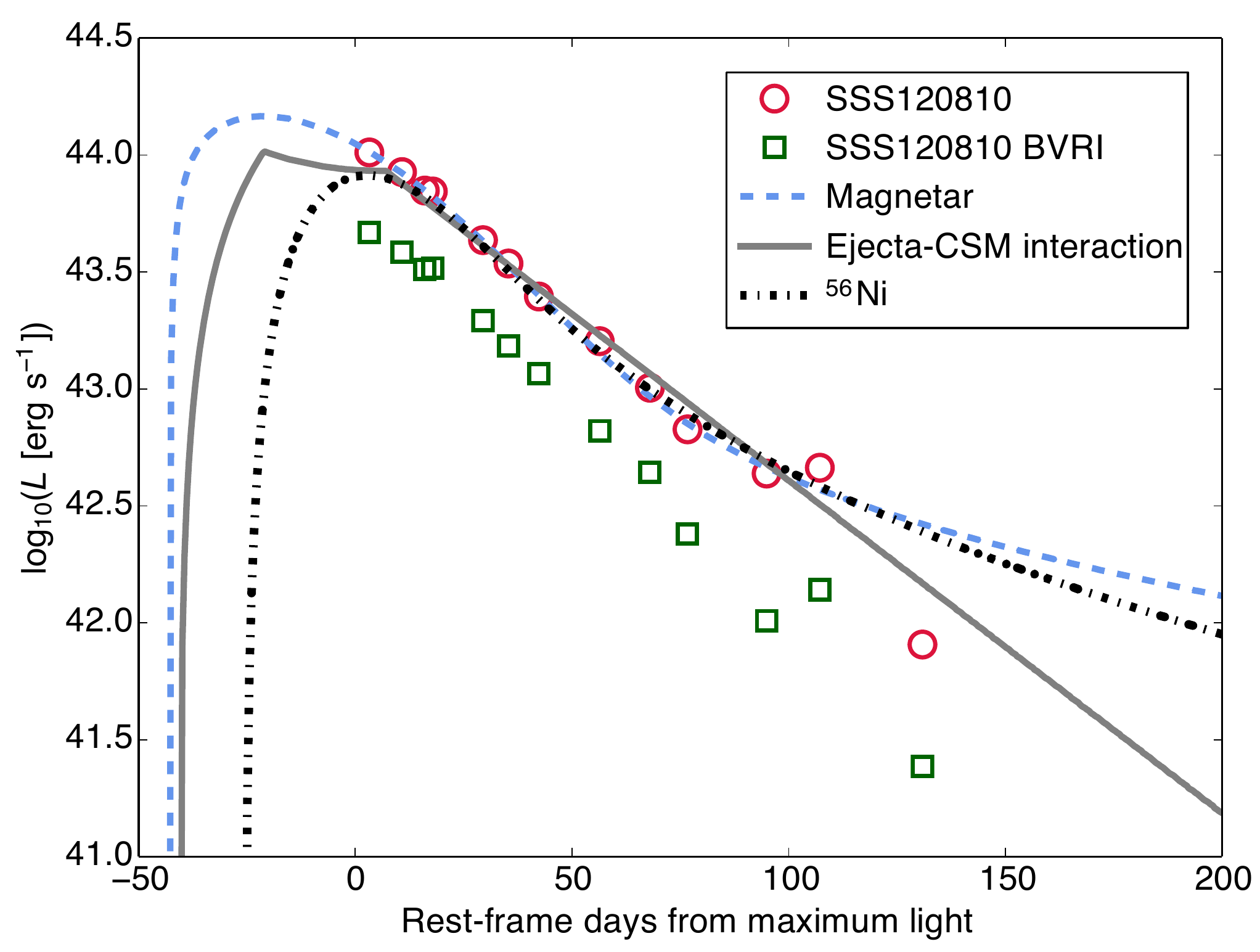}
		}\\
	\subfigure{
		\includegraphics[width=8cm,angle=0]{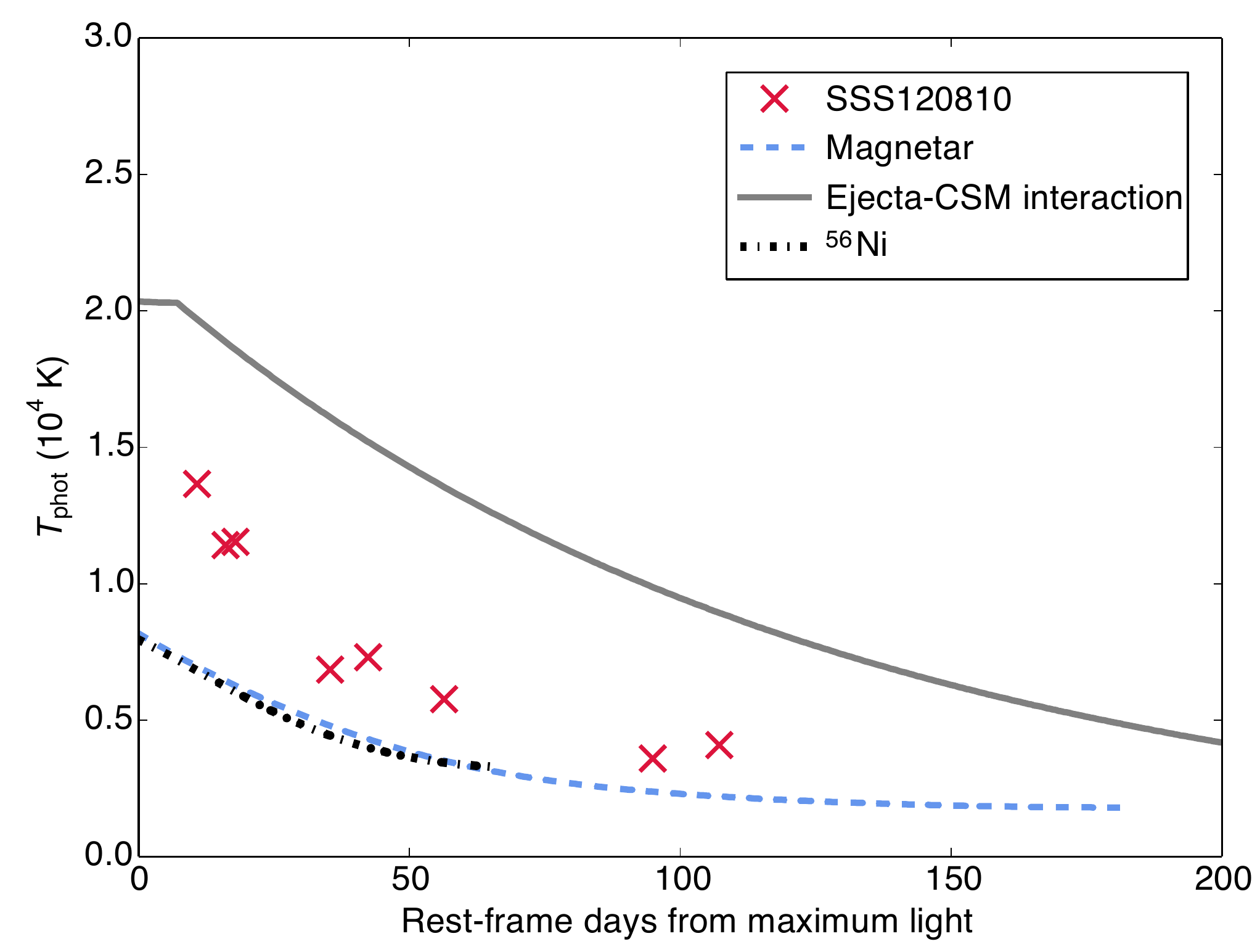}
		}\\
	\subfigure{
		\includegraphics[width=8cm,angle=0]{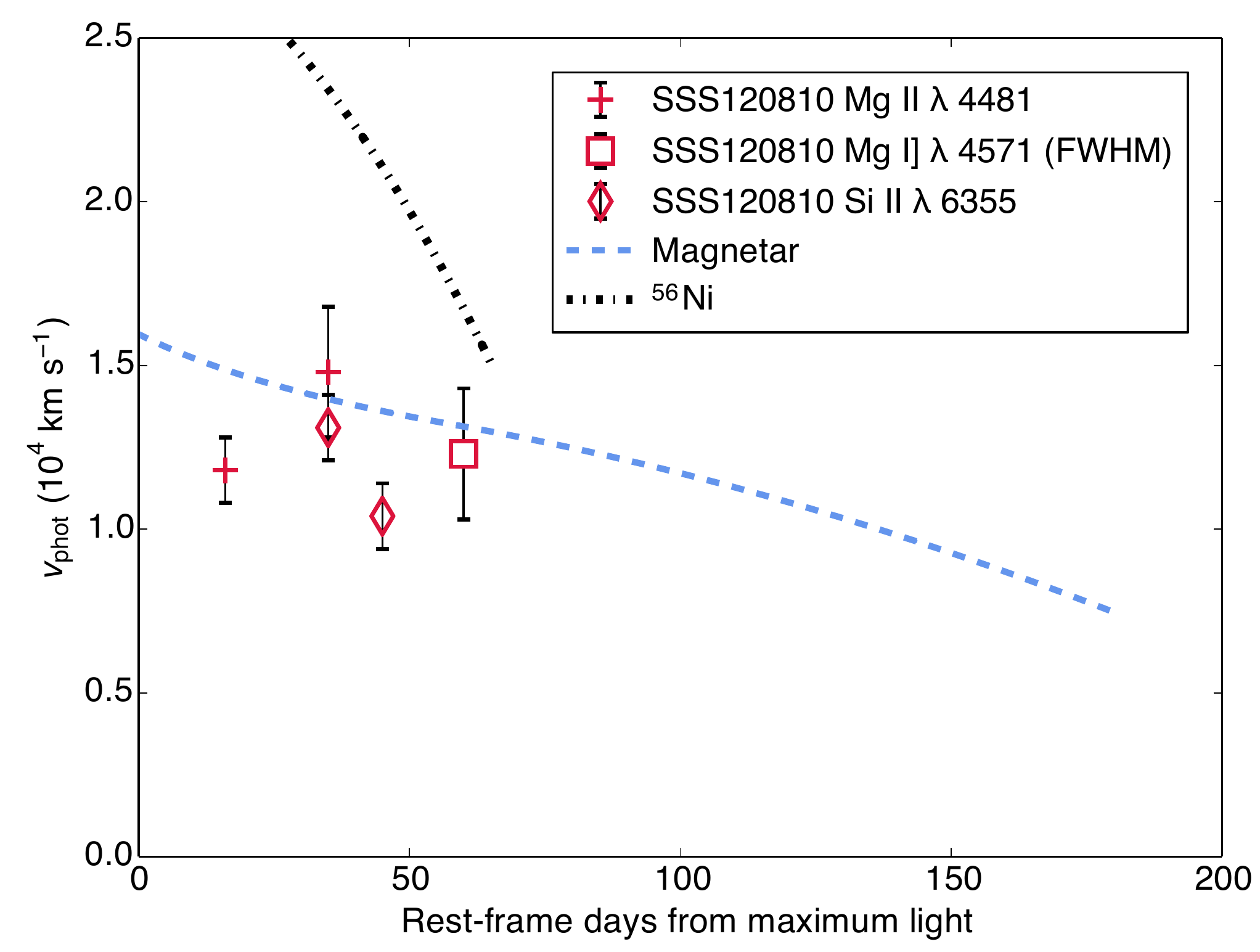}
		}\\
\end{center}
\caption{Magnetar-, interaction- and \Ni-powered models of SSS120810. Parameters are listed in Table \ref{fit_params}. Temperatures were estimated by fitting blackbody curves to multi-colour photometry; velocities were measured from absorption minima.}
\label{sss_models}
\end{figure}

This SN is the most difficult to fit, despite the lack of photometry before maximum brightness, because none of our simple light curve models naturally accommodate a late rebrightening as observed in the SSS120810 data. In Figure \ref{sss_models}, we show both the bolometric and {\it BVRI} pseudo-bolometric light curve, to illustrate the uncertainty in the relative height of this feature. Regardless, the final point on the light curve falls well below the tails of our magnetar and radioactive fits. The interaction fit parameters are quite different from our other light curves, with \Mej $\sim16$ \M~ and \Mcsm $\sim 2$ \M, whereas the other fits have \Mej $\sim4-8$ \M~and \Mej/\Mcsm $\sim1.4-2.2$. However, the parameters for SSS120810 are by far the least well-constrained, due to the lack of early data. The magnetar model gives a good fit to the velocity, while the temperature is intermediate between the magnetar and interaction models. We again reject the radioactive model, because of the inferred 90\% \Ni~ejecta.

One possible explanation for the bump at 100 days could be circumstellar interaction with multiple shells of material. The peak emitted luminosity of a shocked shell should approximately obey the relation \citep{qui2007,smi2007} $L\sim\frac{1}{2}M_{\rm CSM}v_{\rm phot}^2 / t_{\rm rise}$, where $v_{\rm phot}$ is the photospheric velocity and $t_{\rm rise}$, the rise time, is a typical light curve timescale. Let us assume that the main light curve peak ($L\sim10^{44}\,{\rm erg}\,{\rm s}^{-1}$) is powered by an ejecta-CSM interaction as described by our best-fit model in Figure \ref{sss_models} (\Mej $\sim16$ \M; \Mcsm $\sim2$ \M; $t_{\rm rise}\sim30\,$d). Although our code utilises a simplifying stationary photoshere, in reality the shocked shell is expanding, as momentum must be conserved. If it then encounters further material, another shock, and consequently a rebrightening, may occur, with a luminosity also roughly given by the above expression.

In our case, this anomaly appears to be much faster than the main light curve timescale; the final point on our light curve is consistent with the original decline, suggesting the rebrightening lasts $\la30\,$d. This in turn suggests a much lower CSM mass compared to the first shell (remembering that \Mcsm~is an important factor in setting the light curve width). In this scenario, we might expect the outer shell to be swept up by the inner shell/ejecta without causing the expanding material to decelerate significantly. Since the shell velocity is then similar before and after the second collision, we may write $(M_2/M_1)\sim(L_2/L_1)(t_2/t_1)$. Fitting a straight line to our bolometric light curve, we find that the bump is $\sim2.4\times10^{42}{\rm erg}\,{\rm s}^{-1}$ brighter than the predicted luminosity at this phase, and the rise time is $\ga10\,$d. This gives an estimated mass of $\ga 0.01$ \M~for the outer CSM. Perhaps this is associated with a normal stellar wind, prior to ejection of the dense shell. There are related alternatives to this picture, for example a single CSM shell, but with clumpy structure in the outer layer. If the forward shock encounters such a clump, the change in density may cause a rebrightening. Another possibility is a change in the density gradient towards the outside of the shell. However, it should be noted that in our fit the forward shock breaks out of the shell around peak, long before the rebrightening.

Another intriguing possibility is that we have the first optical observation of magnetar wind breakout in a SLSN. \citet{met2014} predict that, under certain conditions, the ionisation front of the pulsar wind nebula could break out of the ejecta a few months after the optical light curve peak. \citet{lev2013} observed X-ray emission from SCP06F6, one of the first known SLSNe Ic, at just such a phase, but no other SLSNe have been detected in X-rays \citep[limits have been measured by][]{ofe2013}. \citet{met2014} found that the ionisation front is more likely to break out (and to break out earlier) for more energetic magnetars, and in fact our fit to SSS120810 suggests a spin period of 1.2 ms -- close to the maximum allowed rotation rate for neutron stars. Those authors also point out that X-ray breakout may result in an abrupt change in the optical properties of the SN, such as the effective temperature. Our observations indicate that the rebrightening is more pronounced at bluer wavelengths, and our estimates of the blackbody colour temperature shown in Figure \ref{sss_models} seem to support the idea that the rebrightening is associated with a reheating of the ejecta.

\subsection{SN 2013dg}

\begin{figure}
\begin{center}
	\subfigure{
		\includegraphics[width=8cm,angle=0]{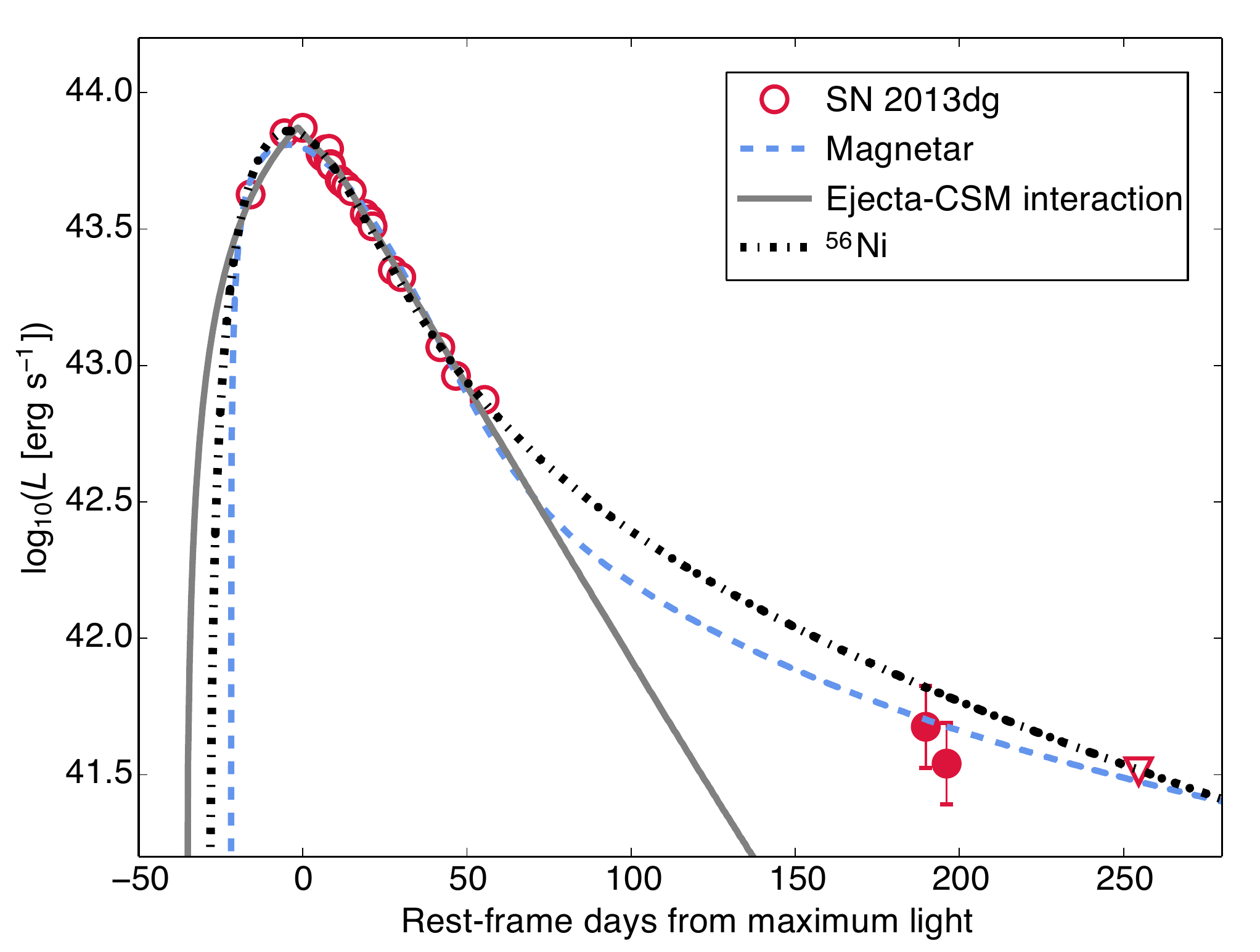}
		}\\
	\subfigure{
		\includegraphics[width=8cm,angle=0]{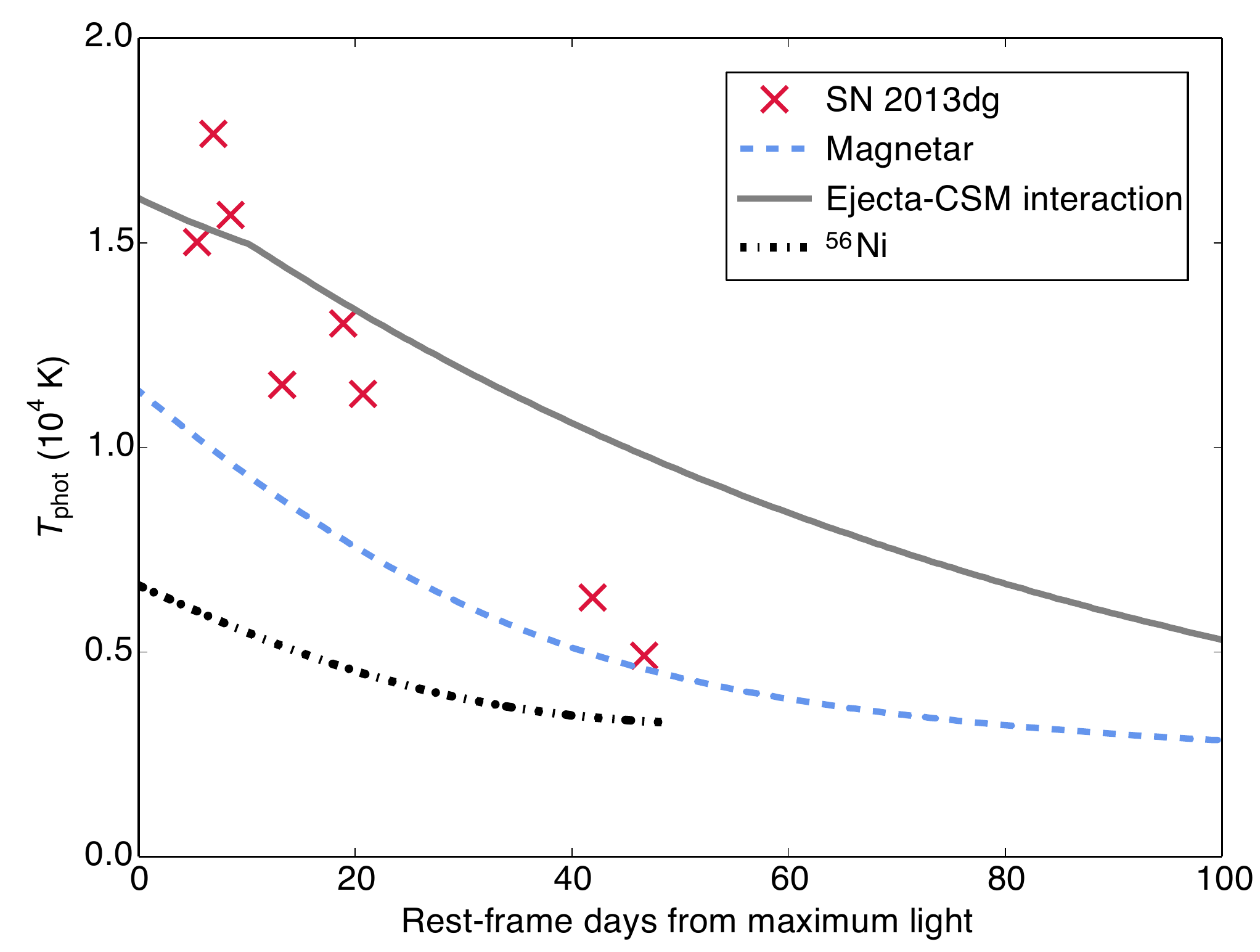}
		}\\
	\subfigure{
		\includegraphics[width=8cm,angle=0]{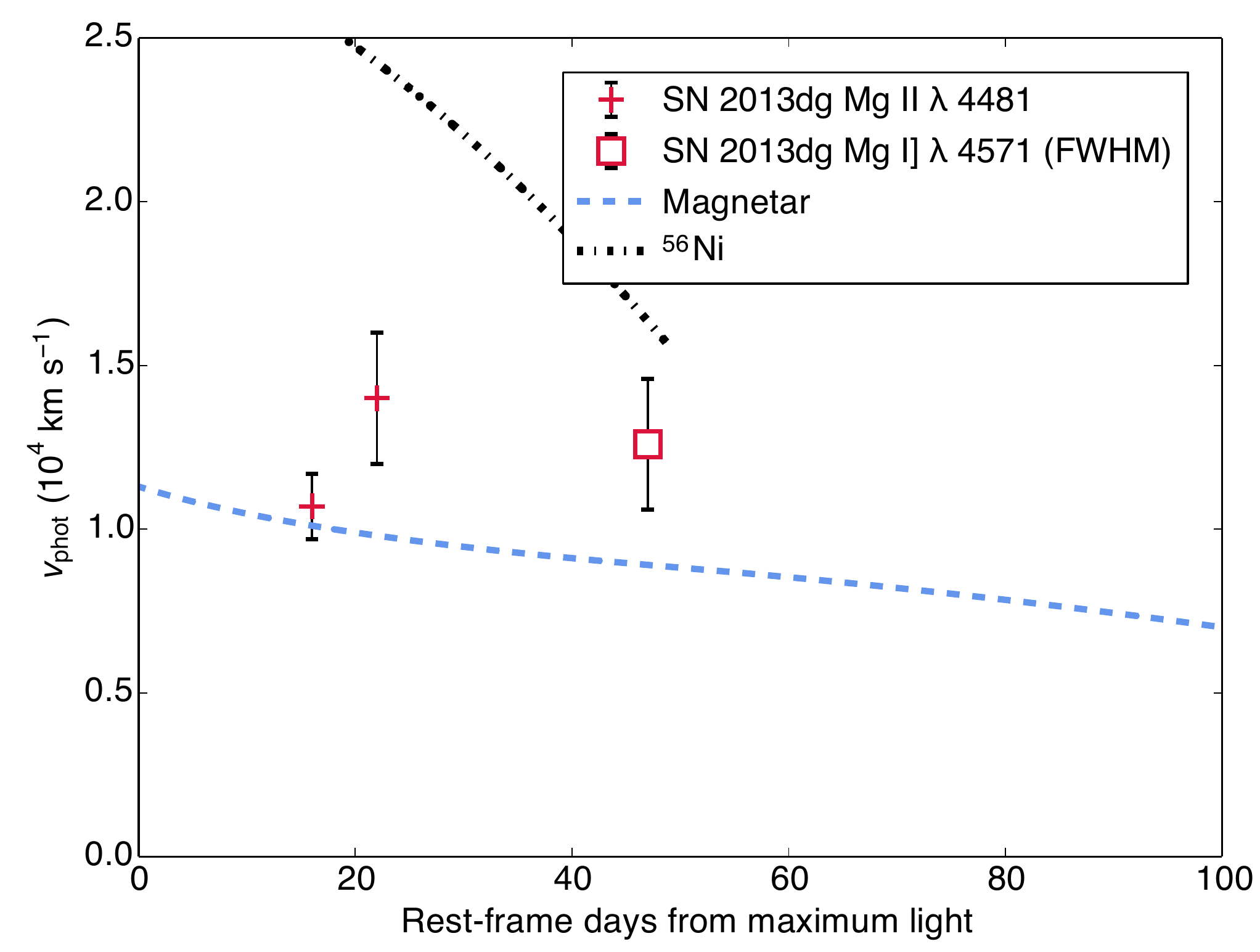}
		}\\
\end{center}
\caption{Magnetar-, interaction- and \Ni-powered models of SN 2013dg. Parameters are listed in Table \ref{fit_params}. Temperatures were estimated by fitting blackbody curves to multi-colour photometry; velocities were measured from absorption minima.}
\label{dg_models}
\end{figure}

\begin{figure}
\begin{center}
	\subfigure{
		\includegraphics[width=8cm,angle=0]{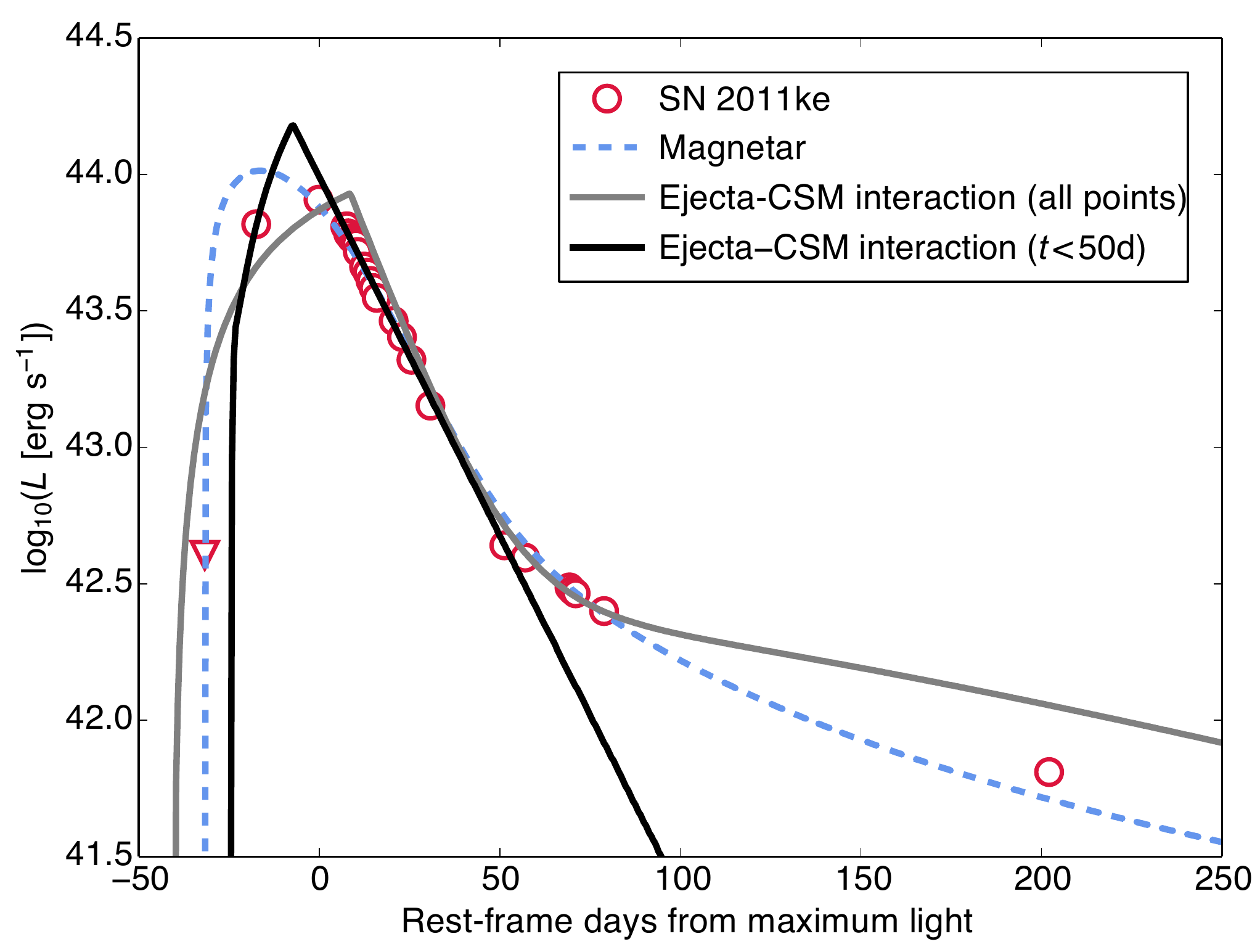}
		}\\
	\subfigure{
		\includegraphics[width=8cm,angle=0]{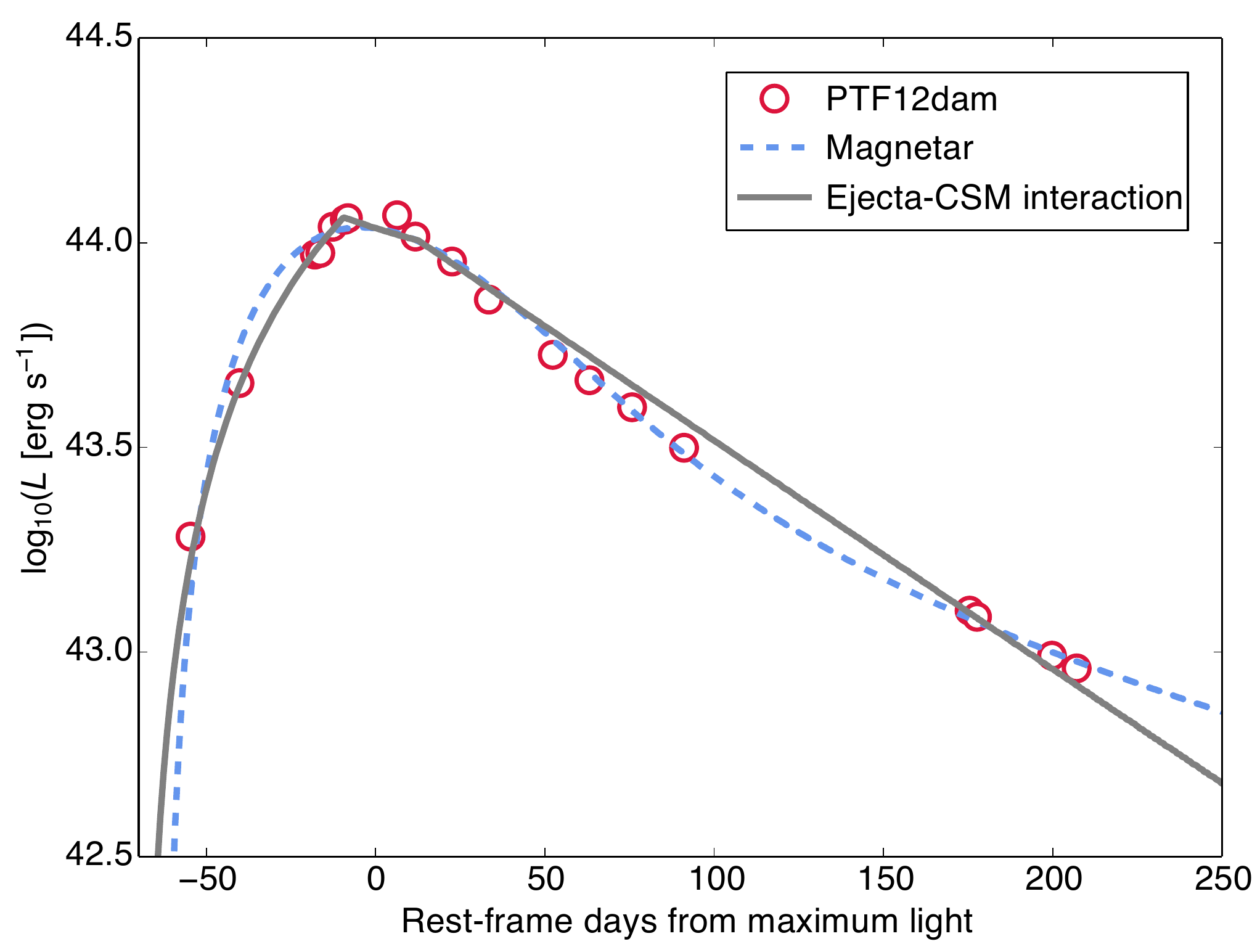}
		}\\
\end{center}
\caption{Magnetar- and interaction-powered models of SN 2011ke and PTF12dam. Magnetar fits are the same as those shown in \citet{ins2013} and \citet{nic2013}. Parameters are listed in Table \ref{fit_params}. \textit{Top}: The light curve of SN 2011ke strongly favours a magnetar. While both models struggle to fit the rise, the magnetar gives a better match to the late tail-phase. In the CSM model, the tail is powered by nickel. The parameters we derive (\Mej $\sim0.8$ \M; \Mni $\sim0.3$ \M; \Mcsm $\sim0.1$ \M) are actually loosely consistent with a SN Ia exploding inside a CSM shell -- a `Ia-CSM' \citep[e.g.][]{silv2013,ald2006,dil2012} \citep[but see also][and references therein]{ins2014}. However, its spectrum is that of a typical SLSN Ic, while all existing Ia-CSM candidates have been hydrogen-rich. To fit the first fifty days of the light curve requires no \Ni, but needs \Mej$\sim11\,$\M~(and far denser CSM). \textit{Bottom}: \citet{nic2013} modelled the slowly declining SLSN Ic PTF12dam and, after ruling out \Ni-powered models, suggested a magnetar model. Our expanded fitting routines now show that an interaction fit is also a viable explanation of its unusual light curve. The derived parameters for PTF12dam are actually similar to the majority of our objects, except that the masses of ejecta and CSM are larger by a factor $\sim5$.}
\label{12dam_11ke}
\end{figure}

As shown in section \ref{phot}, the light curve of SN 2013dg is the most similar to typical low-redshift SLSNe Ic \citep{ins2013}. It is well fit by all of our models (Fig. \ref{dg_models}), although the composition of our radioactive model is typically unrealistic. Points beyond $\sim100$ days are required to distinguish between models, as the magnetar light curve predicts a turn-off not seen in the interaction fit. Our observations at $\sim$200d are in line with the magnetar model; however without a template at a phase of $>$1 year, we do not know how much of this flux comes from the host. For this reason, we exclude these points from the interaction fit, which would require substantial \Ni~mass to replicate this tail ($\ga5$ \M, judging from our radioactive fit here). This amount of \Ni~would have a significant effect on the light curve peak, and our derived parameters. To investigate this, we apply our code to SN 2011ke, which had a light curve very similar to SN 2013dg up to 50 days (Figures \ref{lc_compare} and \ref{bol}), before slowing in its decline. The results are shown in Figure \ref{12dam_11ke}. Fitting the whole 2011ke light curve requires very different parameters to our best-fit model of the peak (and of SN 2013dg) -- essentially we require a SN Ia ejecta and nickel mass, embedded in a fairly low-mass CSM, rather than the massive star model fitting the light curve peak (Table \ref{fit_params}). We note that similar CSM models have been used to explain super-Chandrasekhar mass SNe Ia \citep{tau2011,sca2014}, but that the spectra of such events are very different to SLSNe Ic.

The interaction model does the best job of matching the high early temperatures seen in SN 2013dg, though the SN cools quite quickly, and by 40 days is closer to the magnetar model. The nickel model is rather cool compared to our observations. Measuring the evolution of line velocities proved difficult, as blending meant that few lines were useful at multiple epochs. For example, the maximum of the Fe II feature at $\sim5200$ \AA~moves from 5169 \AA~(the line used for LSQ12dlf) to $\sim5300$ \AA~as the spectrum evolves, artificially inflating the measured velocity with time. Nevertheless, the fairly flat velocity curve of our magnetar fit appears to agree with our estimates for the magnesium layer.

\subsection{CSS121015}

\begin{figure}
\includegraphics[width=8cm,angle=0]{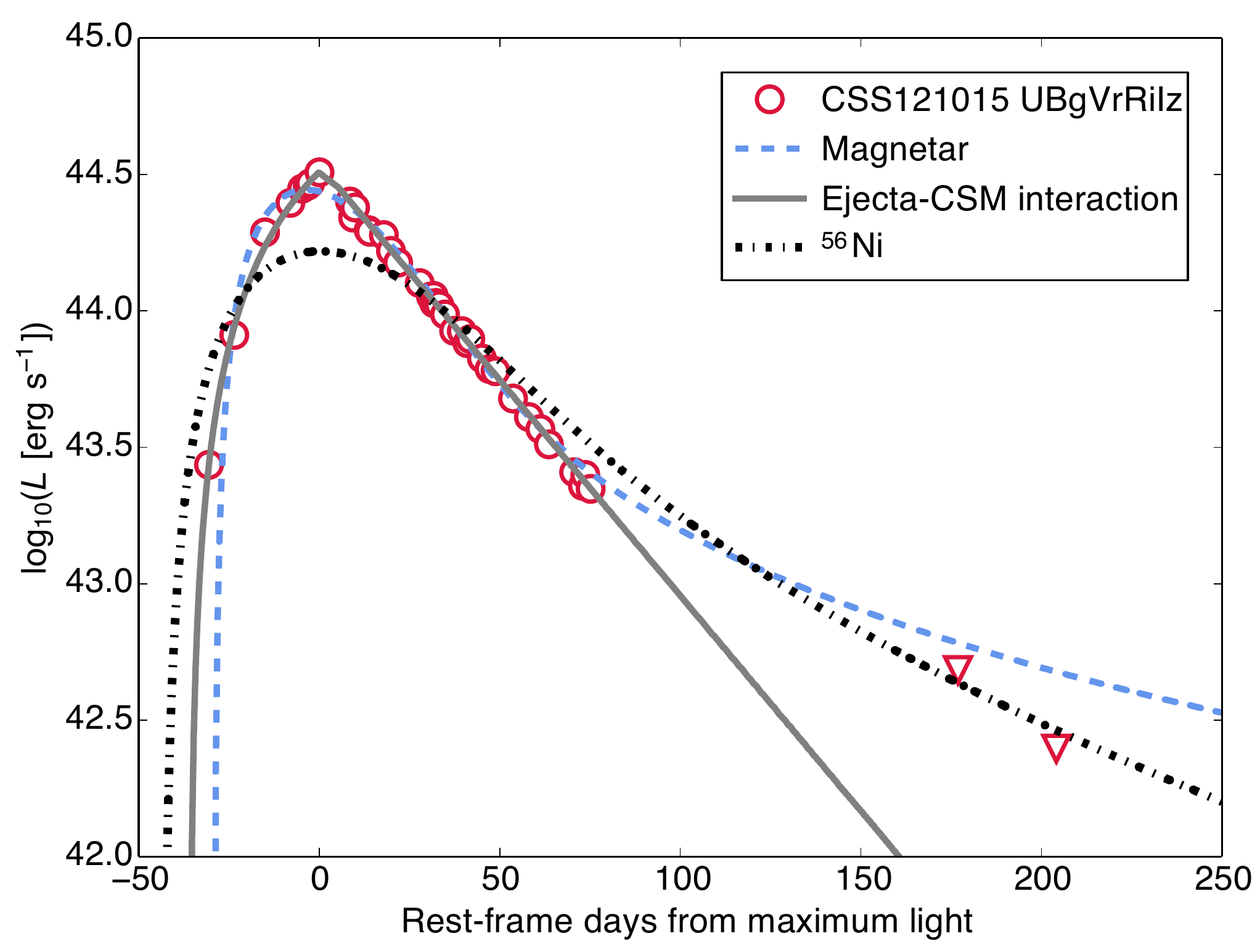}
\caption{Fits to the pseudo-bolometric light curve of CSS121015 \citep{ben2014}, using models powered by magnetar radiation, ejecta-CSM interaction, and \Ni~decay. Parameters are listed in Table \ref{fit_params}.}\label{css_fit}
\end{figure}

The Type II SLSN, CSS121015, was studied by \citet{ben2014}. They found that the observed properties were broadly consistent with a scenario in which the optical transient is powered by the collision of the SN ejecta with several solar masses of CSM. We fit this SN here (Fig. \ref{css_fit}; for parameters see Table \ref{fit_params}) -- both as an extension of that work, and as a test of our model. Our light curve fit supports the interpretation of those authors, with a best-fit \Mcsm $\sim5\,$\M, similar to the $\sim8\,$\M~they estimated. While the quality of fit is similar to that of the magnetar model over the observed lifetime of the SN, the upper limits obtained $\sim$ 200 days after maximum light prove to be useful discriminators. The $t^{-2}$ magnetar tail over-predicts the flux at late epochs, whereas in the interaction model, shock heating terminates a few days after maximum, and the light curve is subsequently just radiative diffusion; the SN therefore lacks a power input to drive a bright tail-phase (though small \Ni~mass cannot be excluded, as a radioactive tail similar to normal core-collapse SNe would not have been detected at this distance). This clearly illustrates the utility of our simple models, and lends support to the arguments presented by \citet{ben2014}. Of course, there is also the possibility that the ejecta no longer traps all of the high-energy magnetar input at late epochs \citep[e.g.][]{kot2013}, causing the optical emission to drop below the prediction of our fully-trapped model.  \Ni~cannot be the dominant power source, as our fit is poor even for very optimistic parameters -- ejecta composed almost entirely of \Ni, with an explosion energy of $10^{53}$erg.

\subsection{CSM configuration}

\begin{figure}
\includegraphics[width=8.5cm,angle=0]{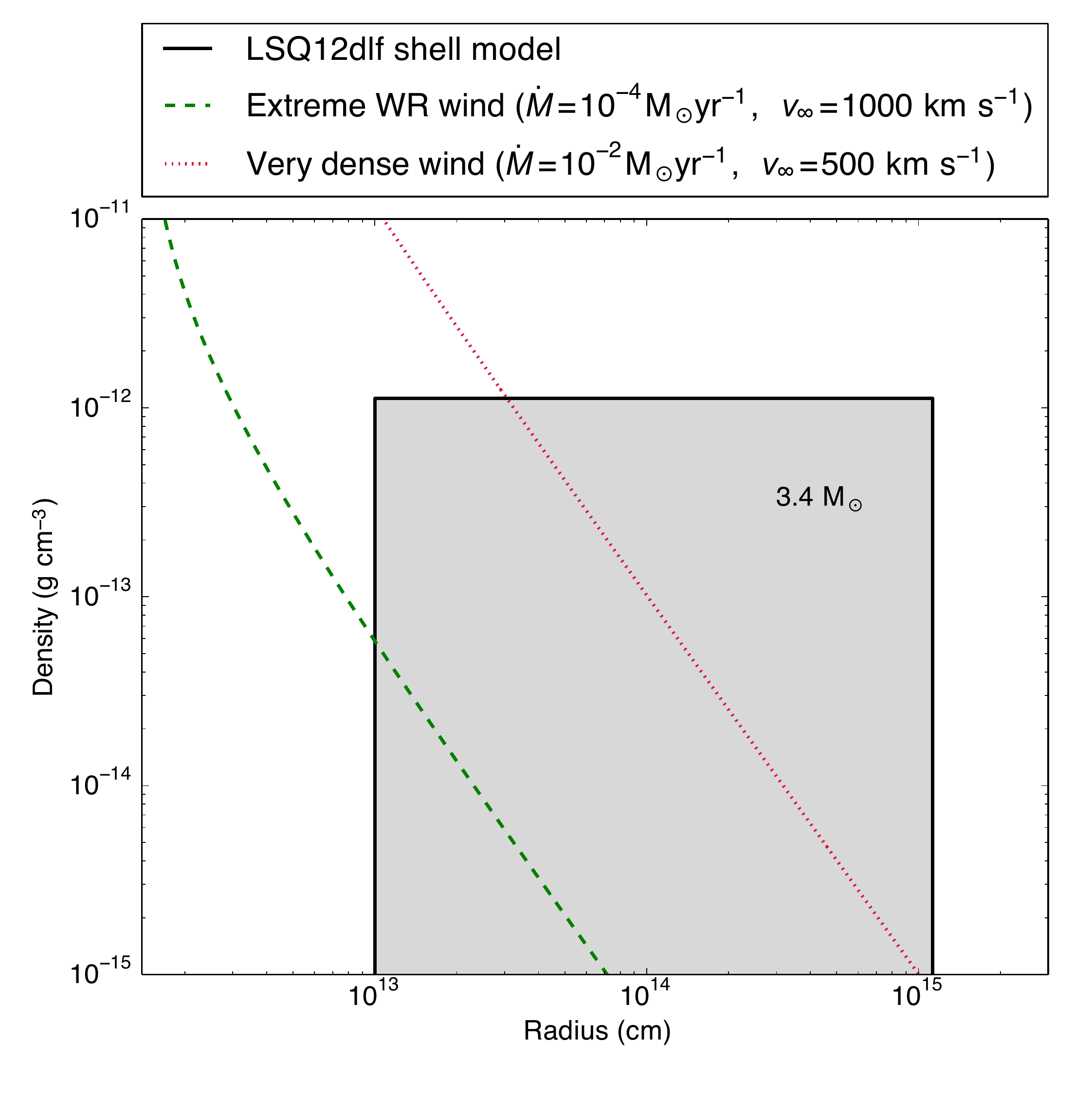}
\caption{The circumstellar density profile used to fit the light curve of LSQ12dlf with our interaction model. Also shown for comparison are a somewhat extreme Wolf-Rayet wind (green dashed line) and a hypothetical dense, slow wind representing enhanced mass-loss shortly before explosion (red dotted line). This demonstrates the difficulty of achieving very high densities at large radii from the progenitor.}\label{csm_config}
\end{figure}

The interaction-powered fits require dense CSM extending to a radius $\sim10^{15}$cm. To put this in context, we compare the model for LSQ12dlf to the densest known winds from Wolf-Rayet (WR) stars. These have mass-loss rates ($\dot{M}$) approaching $10^{-4}\,$\M$\,$yr$^{-1}$, and terminal velocities ($v_{\infty}$) $\sim1000\,$\kms \citep{cro2007,gra2008,hil1999}. This is shown in Figure \ref{csm_config}. To find the density profiles generated by these winds, we use the following parameterisation:
\begin{equation}
   \dot{M}=4 \pi r^2 \rho(r) v(r),
\end{equation}
from the equation of continuity, and
\begin{equation}
   v(r) = v_{\infty}(1-R_{*}/r)^{\beta}
\end{equation}
\citep[see][and references therein]{cro2007}, where $R_{*}$ is the stellar radius. We take $R_{*}=20$R$_{\odot}$ and $\beta=1$ as fiducial values. Values of $\beta$ between 1--5 are typical, and the inferred density profiles are largely insensitive to these choices.

The WR wind falls orders of magnitude short of the densities in our model fit. To get close to the required density, we need rapid mass loss ($\ga10^{-2}\,$\M$\,$yr$^{-1}$) at fairly low velocity ($v_{\infty}\la500\,$\kms), which probably necessitates a massive outburst shortly before the explosion. As pointed out by \citet{gin2012}, the extreme mass-loss needed may help to explain why SLSNe are so rare. \citet{dwa2007} has simulated how SN evolution occurs in circumstellar environments shaped by Wolf-Rayet stars using mass-loss rates and wind velocities typical of WR stars observed in the Local Group. He finds that the optical and X-ray light curves can be significantly affected, but the mass-loss regime explored ($\dot{M} \sim$ few $\times 10^{-5} \, {\rm M_{\odot}} \,$yr$^{-1}$and $v_{\infty} \sim 2000-3000$\kms)
is much lower than we require for the dense shell scenario.

\section{Conclusions}\label{conc}

We have presented light curves and spectra for the three SLSNe Ic classified in the first year of PESSTO. The spectra appear quite homogeneous, and very much in line with other SLSNe Ic, such as SN 2010gx. Little evolution is seen over the post-maximum photospheric phase, during which time the spectra are dominated by broad lines of singly-ionised metals. Despite this similarity, we see a surprising degree of variety in their light curves, with very different decline rates after maximum and, in the case of SSS120810, evidence of a late re-brightening. SN 2013dg shows a possible break in the decline rate at $\sim50\,$d, as previously witnessed in SN 2011ke and others. Such a decline break is not seen in SSS120810 or LSQ12dlf. With these very different declines from maximum, we might expect to see these SNe becoming nebular at different relative phases, so coordinating late-time follow-up will be an important step in understanding the nature of these events.

The light curves were analysed using simple diffusion models with radioactivity, magnetar spin-down and ejecta-CSM interaction as power sources. In developing these models, we followed the work of \citet{ins2013,cha2012,arn1982,che1994}. We find that none of our light curves can be fit with plausible \Ni-powered models. The typical properties of our interaction fits are: \Mej $\,\ga5$ \M; \Mej/\Mcsm $\sim1-2$; $E\sim1-2\times10^{51}\,$erg; \Rph $\sim1-2\times10^{15}\,$cm; $\log(\rho/{\rm g}\,{\rm cm}^{-2})\sim-12$.

For several objects, the late-time evolution appears to be faint compared to magnetar model predictions. However, our models assume full energy trapping at all epochs, whereas in reality this may be a time-dependent process. Our magnetar tails are thus upper-limits to the late-time luminosities of magnetar-powered SNe.

\citet{ins2013} and \citet{nic2013} proposed that magnetar-powered models could explain all of the SLSNe Ic then known, whereas other authors, such as \citet{cha2013} and \citet{ben2014}, favour circumstellar interaction. Our fits here reinforce the validity of both of these interpretations, without particularly favouring either. More detailed hydrodynamical light curve modelling and synthetic spectra are needed to disentangle the signatures of these two possible power sources. In particular, it remains to be seen whether interaction models can reproduce the observed spectra, and if so, under what conditions. If there are two mechanisms at play, we need to understand how these very different processes produce such similar spectra. No SLSN Ic to date has shown narrow lines, the traditional signature of circumstellar interaction (though some SLSNe II, such as CSS121015, showed narrow H lines as well as marked similarity to SLSNe Ic), so the next step is to investigate whether we should expect this to be the case for the regimes of density, temperature and opacity needed to reproduce the light curves \citep[and how this differs from Type Ic SN 2010mb, a lower luminosity pulsational-PISN candidate that did show narrow oxygen lines;][]{ami2014}. Additionally, the presence of broad SN lines in the early spectra may be inconsistent with the presence of an obscuring circumstellar shell. On the observational side, probing the physics of SLSNe will require spectra at very late times, to examine the ejecta composition, and more data in the high-energy regime, to look for signatures of magnetar wind breakout.

\bigskip
\noindent
{\bf ACKNOWLEDGMENTS}
This work is based on observations collected at the European Organisation for Astronomical Research in the Southern Hemisphere, Chile as part of PESSTO (the Public ESO Spectroscopic Survey for Transient Objects), ESO program ID 188.D-3003. VLT+X-shooter spectra were obtained under ESO programs 089.D-0270 and 091.D-0749. Other observations have been collected using: the 4.3m William Herschel Telescope, operated on the island of La Palma by the Isaac Newton Group of Telescope; the Liverpool Telescope, which is operated by Liverpool John Moores University in the Spanish Observatorio del Roque de los Muchachos of the Instituto de Astrofisica de Canarias with financial support from the UK Science and Technology Facilities Council; the Las Cumbres Observatory Global Telescope Network (LCOGTN); the Gemini Observatory, which is operated by the Association of Universities for Research in Astronomy, Inc., under a cooperative agreement with the NSF on behalf of the Gemini partnership: the National Science Foundation (United States), the National Research Council (Canada), CONICYT (Chile), the Australian Research Council (Australia), Minist\'{e}rio da Ci\^{e}ncia, Tecnologia e Inova\c{c}\~{a}o (Brazil) and Ministerio de Ciencia, Tecnolog\'{i}a e Innovaci\'{o}n Productiva (Argentina). Research leading to these results has received funding from the European Research Council under the European Union's Seventh Framework Programme (FP7/2007-2013)/ERC Grant agreement n$^{\rm o}$ [291222] (PI S.J.S). We acknowledge funding from STFC and DEL NI. S.B. is partially supported by the PRIN-INAF 2011 with the project ÓTransient Universe: from ESO Large to PESSTOÓ. M.F. was partly supported by the European Union FP7 programme through ERC grant number 320360. N.E.R. acknowledges the support from the European Union Seventh Framework Programme (FP7/2007-2013) under grant agreement n. 267251 ``Astronomy Fellowships in Italy'' (AstroFIt). A.G.-Y. is supported by ``The Quantum UniverseÓ I-Core program by the Israeli Committee for planning and funding and the ISF, a GIF grant, and the Kimmel award.

%%%% REFERENCES
\bibliographystyle{mn2e}
\bibliography{/Users/matt/Documents/Papers/bib_lib}

%%%% Spectroscopy tables

\begin{table*}
\caption{Spectroscopic observations of LSQ12dlf} \label{lsq_spec}
\begin{tabular}{cccccc}
\hline
\hline
date   & MJD    & phase*  &  Setup &   range (\AA)    & Resolution (\AA)\\
\hline			    		                    
2012-08-07 &56147.3& +7       & NTT+EFOSC2  & 3700-9200             &  18  \\
2012-08-09 &56149.3& +9       & NTT+EFOSC2    & 3400-10000          &  13  \\
2012-08-18& 56158.3& +16      & NTT+EFOSC2      &  3700-9200          &    18       \\
2012-08-24&56164.3& +21      & NTT+EFOSC2     & 3700-9200          &  18  \\
2012-09-09&56180.3& +34      & NTT+EFOSC2        & 3700-9200      &  18     \\
2012-09-12&56182.5& +36     & VLT+Xshooter    & 3100-24000            & 1  \\
2012-09-22&56193.3& +44      & NTT+EFOSC2        & 3700-9200      &  18     \\
2012-12-09&56270.5& +106      & Gemini S.+GMOS    & 4660-8900         & 2     \\
\hline
\end{tabular}

* Phase in rest-frame days relative to epoch of maximum light.

\end{table*}

\begin{table*}
\caption{Spectroscopic observations of SSS120810} \label{sss_spec}
\begin{tabular}{cccccc}
\hline
\hline
date   & MJD    & phase*  &  Setup &   range (\AA)    & Resolution (\AA)\\
\hline			    		                    
2012-08-17 &56158.3& +10      & NTT+EFOSC2      & 3700--9200         &  18  \\
2012-08-18& 56158.3& +11       & NTT+EFOSC2      &  3700--9200         &    18       \\
2012-08-24&56164.3& +16       & NTT+EFOSC2     & 3700--9200         &  18  \\
2012-09-15&56186.2 & +35      & NTT+EFOSC2      & 3700--9200        &  18     \\
2012-09-23&56194.2& +42     &NTT+EFOSC2        & 3700--9200      & 18  \\
2012-09-26&56196.5& +44      & VLT+Xshooter     &  3100--24000         & 1     \\
2012-10-14&56215.3& +60        & NTT+EFOSC2  & 3700--9200           & 18     \\

\hline
\end{tabular}

* Phase in rest-frame days relative to epoch of maximum light.

\end{table*}

\begin{table*}
\caption{Spectroscopic observations of SN 2013dg} \label{13dg_spec}
\begin{tabular}{cccccc}
\hline
\hline
date   & MJD    & phase*  &  Setup &   range (\AA)    & Resolution (\AA)\\
\hline			    		                    
2013-06-10 &56454.0& +4      &WHT+ISIS      & 3260--10000         &  $4.1-7.7$  \\
2013-06-13& 56457.0& +6        & Gemini S.+GMOS      &  4660--8900         &   2       \\
2013-06-25&56469.0& +16       &  VLT+Xshooter      & 3100--24000         &  1  \\
2013-07-03&56477.0 & +22       & Gemini S.+GMOS       & 4660--8900      &  2     \\
2013-07-20&56493.0& +35     & VLT+Xshooter        & 3100--24000       & 1  \\
2012-08-03&56508.0& +47       &Gemini S.+GMOS     &  4660--8900       & 2     \\

\hline
\end{tabular}

* Phase in rest-frame days relative to epoch of maximum light.

\end{table*}

%%%% Photometry tables

\begin{table*}
\caption{Observed photometry of LSQ12dlf} \label{lsq_phot_tab}
\begin{flushleft}
\begin{tabular}{ccccccccc}
\hline
\hline
Date & MJD & Phase$^*$ & U & B & V & R & I & Instrument$^{**}$\\
\hline			    		                    
2012-06-18 & 56097.41 & -32.7 &  &  & \textgreater22.32  &  &  & LSQ \\
2012-06-18 & 56097.44 & -32.6 &  &  & \textgreater21.93  &  &  & LSQ \\
2012-06-22 & 56101.36 & -29.5 &  &  & \textgreater22.08  &  &  & LSQ \\
2012-06-22 & 56101.41 & -29.5 &  &  & \textgreater22.17  &  &  & LSQ \\
2012-07-09 & 56118.35 & -15.9 &  &  & 19.33 (0.03) &  &   & LSQ \\
2012-07-09 & 56118.41 & -15.9 &  &  & 19.25 (0.02) &  &   & LSQ \\
2012-07-15 & 56124.41 & -11.1 &  &  & 19.07 (0.02) &  &   & LSQ \\
2012-07-15 & 56124.43 & -11.1 &  &  & 19.02 (0.02) &  &   & LSQ \\
2012-07-17 & 56126.41 & -9.5 &  &  & 19.04 (0.03) &  &   & LSQ \\
2012-07-17 & 56126.43 & -9.5 &  &  & 19.06 (0.03) &  &   & LSQ \\
2012-07-21 & 56130.31 & -6.4 &  &  & 18.89 (0.03) &  &   & LSQ \\
2012-07-21 & 56130.38 & -6.3 &  &  & 18.92 (0.03) &  &   & LSQ \\
2012-07-23 & 56132.41 & -4.7 &  &  & 18.92 (0.02) &  &   & LSQ \\
2012-07-23 & 56132.42 & -4.7 &  &  & 18.91 (0.02) &  &   & LSQ \\
2012-07-27 & 56136.24 & -1.6 &  &  & 18.87 (0.04) &  &   & LSQ \\
2012-07-27 & 56136.32 & -1.5 &  &  & 18.90 (0.02) &  &   & LSQ \\
2012-07-29 & 56138.25 & 0.0 &  &  & 18.78 (0.04) &  &   & LSQ \\
2012-07-29 & 56138.34 & 0.1 &  &  & 18.80 (0.03) &  &   & LSQ \\
2012-08-09 & 56149.40 &  8.9 & 18.46 (0.09) & 19.43 (0.04) & 19.15 (0.04) & 19.01 (0.09) & 18.82 (0.10) & NTT \\
2012-08-12 & 56152.10 &  11.1 &  & 19.45 (0.23) & 19.22 (0.14) & 19.09 (0.15) & 18.93 (0.12) & LT \\
2012-08-18 & 56158.10 &  15.9 &  & 19.71 (0.13) & 19.22 (0.10) & 18.96 (0.06) & 18.91 (0.05) & LT \\
2012-08-18 & 56158.30 &  16.0 & 18.91 (0.09) & 19.63 (0.06) & 19.26 (0.07) & 19.04 (0.05) & 18.83 (0.17) & NTT \\
2012-08-18 & 56158.40 & 16.1 &  &  & 19.29 (0.02) &  &   & LSQ \\
2012-08-18 & 56158.41 & 16.1 &  &  & 19.39 (0.03) &  &   & LSQ \\
2012-08-20 & 56160.31 & 17.6 &  &  & 19.39 (0.08) &  &   & LSQ \\
2012-08-20 & 56160.39 & 17.7 &  &  & 19.44 (0.03) &  &   & LSQ \\
2012-08-24 & 56164.30 &  20.8 & 19.34 (0.07) & 19.90 (0.04) & 19.42 (0.03) & 19.20 (0.04) & 18.88 (0.05) & NTT \\
2012-08-24 & 56164.35 & 20.9 &  &  & 19.49 (0.03) &  &   & LSQ \\
2012-08-24 & 56164.38 & 20.9 &  &  & 19.60 (0.04) &  &   & LSQ \\
2012-08-26 & 56166.21 & 22.4 &  &  & 19.53 (0.04) &  &   & LSQ \\
2012-08-26 & 56166.29 & 22.4 &  &  & 19.45 (0.03) &  &   & LSQ \\
2012-08-26 & 56166.40 &  22.5 & 19.48 (0.03) & 19.99 (0.03) & 19.59 (0.04) & 19.27 (0.04) & 19.00 (0.04) & NTT \\
2012-09-09 & 56180.40 &  33.7 & 20.31 (0.34) & 20.69 (0.08) & 20.07 (0.06) & 19.59 (0.06) & 19.13 (0.09) & NTT \\
2012-09-12 & 56183.10 &  35.9 &  & 20.86 (0.15) & 20.00 (0.05) & 19.50 (0.07) & 19.43 (0.10) & LT \\
2012-09-15 & 56186.30 &  38.4 &  & 21.05 (0.03) & 20.24 (0.02) & 19.79 (0.03) & 19.45 (0.04) & NTT \\
2012-09-19 & 56190.10 &  41.5 &  & 21.28 (0.18) & 20.30 (0.09) & 19.82 (0.06) & 19.57 (0.13) & LT \\
2012-09-22 & 56193.40 &  44.1 &  & 21.62 (0.07) & 20.46 (0.07) & 19.92 (0.06) & 19.56 (0.05) & NTT \\
2012-09-24 & 56195.00 &  45.4 &  &  & 20.49 (0.31) & 19.98 (0.16) & 19.64 (0.16) & LT \\
2012-10-04 & 56205.10 &  53.5 &  &  & 20.66 (0.30) & 20.12 (0.10) & 20.01 (0.13) & LT \\
2012-10-06 & 56207.40 &  55.3 &  &  & 20.79 (0.11) & 20.11 (0.15) & 19.63 (0.24) & NTT \\
2012-10-12 & 56213.00 &  59.8 &  &  & 21.20 (0.20) & 20.44 (0.20) & 20.01 (0.18) & LT \\
2012-10-20 & 56221.00 &  66.2 &  & 22.64 (0.31) & 21.16 (0.19) & 20.76 (0.18) & 20.45 (0.14) & LT \\
2012-11-06 & 56238.20 &  80.0 &  & 23.32 (0.07) & 21.92 (0.06) & 21.18 (0.05) & 20.59 (0.04) & NTT \\
2012-12-05 & 56267.20 &  103.2 &  & 24.43 (0.33) & 22.82 (0.09) & 22.17 (0.08) & 21.72 (0.51) & NTT \\
2012-01-04 & 56297.00 &  127.2 &  &  &  & 22.71 (0.30) &  & NTT \\
2012-01-11 & 56304.10 &  132.7 &  &  & 24.00 (0.14) &  &   & NTT \\
\hline
2013-10-09 (host) & 56575.3 & 354.8 &  &  & 25.02 (0.15) & $>23.98$ & $>22.65$ & NTT\\
2014 stack (host)$\ddagger$ &  &  &  &  & 24.81 (0.34) &  &  & NTT\\
\hline
\end{tabular}

* Phase in rest-frame days from epoch of maximum light

** LSQ = La Silla QUEST survey\\ NTT = ESO NTT + EFOSC2 (PESSTO)\\ LT = Liverpool Telescope + RATCam
 
 $\ddagger$ Sum of images obtained on 4 nights in Jan and Feb 2014
 
\end{flushleft}
\end{table*}

\begin{table*}
\caption{Observed photometry of SSS120810} \label{sss_phot_tab}
\begin{flushleft}
\begin{tabular}{ccccccccc}
\hline
\hline
Date & MJD & Phase$^*$ & U & B & V & R & I & Instrument$^{**}$\\
\hline			    		                    
2012-08-10 & 56149.7 &3.2  &  &  &  & 17.76 (0.11) &  & SSS\\
2012-08-18 & 56158.3 &10.6 & 17.38 (0.03) & 18.16 (0.03) & 18.09 (0.02) & 17.97 (0.02) & 17.82 (0.03) & NTT\\
2012-08-24 & 56164.3 &15.8 & 17.68 (0.01) & 18.42 (0.01) & 18.23 (0.02) & 18.11 (0.02) & 17.90 (0.03) & NTT\\
2012-08-26 & 56166.4 &17.6 & 17.80 (0.01) & 18.44 (0.02) & 18.22 (0.02) & 18.12 (0.02) & 17.93 (0.02) & NTT\\
2012-09-08 & 56179.6 &29.1  &  &  &  & 18.57 (0.09) &  & SSS\\
2012-09-15 & 56186.3 &34.9 & 19.92 (0.03) & 19.99 (0.01) & 19.21 (0.02) & 18.80 (0.02) & 18.42 (0.02) & NTT\\
2012-09-23 & 56194.3 &41.8 & 20.63 (0.12) & 20.56 (0.03) & 19.62 (0.02) & 19.14 (0.03) & 18.70 (0.04) & NTT\\
2012-10-09 & 56210.3 &55.6  &  & 21.26 (0.14) & 20.27 (0.06) & 19.69 (0.06) & 19.14 (0.17) & NTT\\
2012-10-23 & 56223.5 &67.0  &  &  & 20.78 (0.30) & 20.12 (0.30) & 19.58 (0.30) & FTS\\
2012-11-02 & 56233.4 &75.6  &  &  &  & 20.85 (0.24) &  & FTS\\
2012-11-22$\ddagger$ & 56254.2 &93.6  &  & 24.62 (0.50) & 23.16 (0.13) & 21.57 (0.07) & 20.88 (0.07) & NTT\\
2012-12-06$\ddagger$ & 56268.2 &105.7  &  & 24.20 (0.38) & 22.33 (0.08) & 21.30 (0.07) & 20.72 (0.10) & NTT\\
2013-01-02$\ddagger$ & 56295.2 &129.0  &  & $>25.39$ & $>25.06$ & 23.19 (0.18) &  & NTT\\
\hline
2013-10-09 (host) & 56575.1 & 371.2 &   &   &   & 21.89 (0.07) & 21.14 (0.07) & NTT\\
2013-10-24 (host) & 56590.0 & 384.1 &   & 22.91 (0.06) & 22.27 (0.04) &   &   & NTT\\
\hline
\end{tabular}

* Phase in rest-frame days from estimated epoch of maximum light

** SSS = Siding Springs Survey (CRTS)\\ NTT = ESO NTT + EFOSC2 (PESSTO)\\ FTS = Faulkes Telescope South + Faulkes Spectral 01
 
$\ddagger$ Magnitudes measured after subtracting host images from Oct 2013

\end{flushleft}
\end{table*}

\begin{table*}
\caption{Observed photometry of SN 2013dg} \label{dg_phot_tab}
\begin{flushleft}
\begin{tabular}{cccccccc}
\hline
\hline
Date & MJD & Phase$^*$ & $g$& $r$ & $i$ & $z$ & Instrument$^{**}$\\
\hline			    		                    
%2013-04-29 & 56411.3 & -30.1 &  & \textgreater21.7  &  &  & CSS \\
2013-05-13 & 56425.2 & -19.0 &  & 20.27 (0.54) &  &  & CSS \\
2013-05-17 & 56429.2 & -15.9 &  & 19.72 (0.17) &  &  & MLS \\
2013-05-30 & 56442.2 & -5.6 &  & 19.16 (0.23) &  &  & CSS \\
%2013-06-01 & 56444.2 & -4.0 &  & 19.18 (0.03) &  &  & MLS \\
2013-06-06 & 56449.2 & 0.0 &  & 19.11 (0.17) &  &  & CSS \\
2013-06-12 & 56456.0 & 5.4 & 19.26 (0.03) & 19.31 (0.06) & 19.50 (0.07) & 19.60 (0.07) & LT \\
%2013-06-13 & 56456.2 & 5.6 &  & 19.33 (0.05) &  &  & MLS \\
2013-06-13 & 56456.9 & 6.1 & 19.36 (0.03) & 19.31 (0.06) & 19.56 (0.10) & & LCO \\
2013-06-14 & 56457.9 & 6.9 & 19.26 (0.07) & 19.36 (0.04) & 19.39 (0.08) & 19.40 (0.17) & LT \\
2013-06-15 & 56459.1 & 7.9 & 19.41 (0.03) & 19.36 (0.07) &  &  & LCO \\
2013-06-15 & 56459.3 & 8.0 &  &  & 19.47 (0.08) & 19.44 (0.22) & FTN \\
2013-06-16 & 56459.9 & 8.5 & 19.42 (0.10) & 19.39 (0.05) & 19.56 (0.10) & 19.69 (0.15) & LT \\
2013-06-16 & 56460.1 & 8.7 & 19.48 (0.06) & 19.34 (0.05) &  &  & LCO \\
2013-06-19 & 56463.4 & 11.3 &  & 19.52 (0.17) &  &  & FTN \\
2013-06-20 & 56464.1 & 11.8 & 19.57 (0.05) & 19.55 (0.05) & 19.68 (0.12) & & LCO \\
2013-06-22 & 56465.9 & 13.3 & 19.68 (0.10) & 19.52 (0.09) & 19.71 (0.13) & 19.62 (0.20) & LT \\
2013-06-25 & 56468.0 & 14.9 & 19.81 (0.05) & 19.54 (0.05) & 19.61 (0.05) & & LCO\\
2013-06-29 & 56473.0 & 18.9 & 20.03 (0.03) & 19.74 (0.03) & 20.04 (0.22) & 19.74 (0.19) & LT \\
2013-07-01 & 56475.0 & 20.5 & 20.20 (0.03) & 19.81 (0.04) & 19.93 (0.06) & & LCO \\
2013-07-01 & 56475.3 & 20.7 & 20.25 (0.17) & 19.80 (0.14) & 19.84 (0.08) & 19.92 (0.18) & FTN \\
2013-07-02 & 56476.0 & 21.3 & 20.26 (0.04) & 19.87 (0.04) & 20.02 (0.08) & & LCO \\
2013-07-10 & 56483.8 & 27.5 & 20.71 (0.08) & 20.23 (0.08) & 20.27 (0.11) & & LCO \\
2013-07-13 & 56487.0 & 30.0 & 20.95 (0.07) & 20.23 (0.06) & 20.24 (0.11) & & FTN \\
2013-07-29 & 56502.0 & 41.9 & 21.48 (0.23) & 20.82 (0.39) & 20.88 (0.22) & 20.52 (0.30) & NTT \\
2013-08-03 & 56508.0 & 46.7 & 22.23 (0.05) & 21.21 (0.04) & 20.95 (0.06) & 20.79 (0.10) & NTT \\
2013-08-14 & 56519.0 & 55.4 &  & 21.88 (0.07) &  &  & NTT \\
2013-08-15 & 56520.0 & 56.2 &  &  & 21.62 (0.16) &  & NTT \\
2013-08-16 & 56520.5 & 56.6 &  &  &  & 20.91 (0.30) & NTT \\
2014-01-30$\ddagger$ & 56688.3 & 189.8 &  &  25.21 (0.41) & 24.68 (0.28) &  & NTT\\
2014-02-07$\ddagger$ & 56696.3 & 196.1 &  &  25.48 (0.36) & 25.21 (0.36) &  & NTT\\
\hline
2014-04-22 & 56770.0 & 254.6 &  &  $>25.63$ & $>25.06$ &  & NTT\\
\hline
\end{tabular}

* Phase in rest-frame days from epoch of maximum light

** CSS = Catalina Sky Survey Survey (CRTS)\\MLS = Mt. Lemmon Survey (CRTS)\\ LT = Liverpool Telescope + RATCam\\LCO = Las Cumbres Observatory Global Telesope 1m Network\\ FTN = Faulkes Telescope North + Faulkes Spectral 02\\ NTT = ESO NTT + EFOSC2 (PESSTO)\\
 
$\ddagger$ May include significant host contribution
\end{flushleft}
\end{table*}

%%%% End of tables

\end{document}